\documentclass[aps,prb,superscriptaddress,floatfix,amsmath,amsfonts,amssymb,showpacs,reprint]{revtex4-2}
\usepackage[utf8]{inputenc}
\usepackage{pifont}
\usepackage{csquotes}
\usepackage[pdftex]{graphicx}    
\usepackage{xcolor}
\usepackage{soul} 
\usepackage{multirow}
\usepackage{bm}
\usepackage[sectionbib]{bibunits}
\defaultbibliographystyle{revtex4-2}

\usepackage[hypertexnames=false]{hyperref}
\definecolor{dark-red}{rgb}{0.9,0.15,0.15}
\definecolor{dark-blue}{rgb}{0.15,0.15,0.4}
\definecolor{medium-blue}{rgb}{0,0,0.5}
\hypersetup{
    colorlinks={true},
    linkcolor={medium-blue},
    citecolor={blue},
    urlcolor={medium-blue},
    pdfproducer={},
    pdfcreator={}
}

\makeatletter
\def\maketitle{
\@author@finish
\title@column\titleblock@produce
\suppressfloats[t]}
\makeatother

\begin{document}

\title{GdAlSi: An antiferromagnetic topological Weyl semimetal with non-relativistic spin splitting}

\author{Jadupati Nag}
\thanks{These authors contributed equally to this work.}
\affiliation{Department of Physics, Indian Institute of Technology Bombay, Mumbai 400076, India}
\affiliation{Graduate School of Advanced Science and Engineering, Hiroshima University, 1-3-1 Kagamiyama, Higashi-Hiroshima 739-8526, Japan}
\affiliation{Dept. of Materials Science and Engineering, \& Materials Research Institute 
Pennsylvania State University, Millennium Sciences Complex Building
University Park, PA 16802, USA}

\author{Bishal Das}
\thanks{These authors contributed equally to this work.}
\affiliation{Department of Physics, Indian Institute of Technology Bombay, Mumbai 400076, India}

\author{Sayantika Bhowal} 
\affiliation{Materials Theory, ETH Zurich, Wolfgang-Pauli-Strasse 27, 8093 Zurich, Switzerland} 

\author{Yukimi Nishioka}
\affiliation{Graduate School of Advanced Science and Engineering, Hiroshima University, 1-3-1 Kagamiyama, Higashi-Hiroshima 739-8526, Japan}

\author{Barnabha Bandyopadhyay}
\affiliation{Department of Physics, Indian Institute of Technology Bombay, Mumbai 400076, India}

\author{Saugata Sarker} 
\affiliation{Dept. of Materials Science and Engineering, \& Materials Research Institute 
Pennsylvania State University, Millennium Sciences Complex Building
University Park, PA 16802, USA}

\author{Shiv Kumar}
\affiliation{Hiroshima Synchrotron Radiation Center (HiSOR), Hiroshima University, 2-313 Kagamiyama, Higashi-Hiroshima 739-0046, Japan}

\author{Kenta Kuroda}
\affiliation{Graduate School of Advanced Science and Engineering, Hiroshima University, 1-3-1 Kagamiyama, Higashi-Hiroshima 739-8526, Japan}
\affiliation{International Institute for Sustainability with Knotted Chiral Meta Matter (WPI-SKCM$^2$), 1-3-1 Kagamiyama, Higashi-Hiroshima 739-8526, Japan}

\author{Venkatraman Gopalan}
\affiliation{Dept. of Materials Science and Engineering, \& Materials Research Institute 
Pennsylvania State University, Millennium Sciences Complex Building
University Park, PA 16802, USA}

\author{Akio Kimura}
\email{akiok@hiroshima-u.ac.jp}
\affiliation{Graduate School of Advanced Science and Engineering, Hiroshima University, 1-3-1 Kagamiyama, Higashi-Hiroshima 739-8526, Japan}
\affiliation{International Institute for Sustainability with Knotted Chiral Meta Matter (WPI-SKCM$^2$), 1-3-1 Kagamiyama, Higashi-Hiroshima 739-8526, Japan}

\author{K. G. Suresh}
\email{suresh@phy.iitb.ac.in}
\affiliation{Department of Physics, Indian Institute of Technology Bombay, Mumbai 400076, India}

\author{Aftab Alam}
\email{aftab@iitb.ac.in}
\affiliation{Department of Physics, Indian Institute of Technology Bombay, Mumbai 400076, India}


\begin{abstract}

 Spintronics has emerged as a viable alternative to traditional electronics based technologies in the past few decades. While on one hand, the discovery of topological phases of matter with protected spin-polarized states has opened up exciting prospects, recent revelation of intriguing non-relativistic spin-splitting in antiferromagnetic (AFM) materials with unique symmetries facilitate a wide possibility of realizing both these features simultaneously. In this work, we report the co-existence of these two intriguing properties within a single material: GdAlSi. Single crystal of GdAlSi stabilizes in a body-centered tetragonal structure with a non-centrosymmetric space group $I4_{1}md$ ($109$), which is confirmed using detailed structural analysis through X-ray diffraction (XRD) and optical second harmonic generation (SHG) measurements. The magnetization data indicates AFM ordering with an ordering temperature ($T_N$) $\sim$32 K. \textit{Ab-initio} calculations reveal GdAlSi to be a collinear AFM Weyl semimetal with an unconventional, momentum-dependent spin-splitting, also referred to as \textit{altermagnet}. Angle-resolved photoemission spectroscopy measurements on GdAlSi single crystals subsequently confirm the presence of Fermi arcs, a distinctive hallmark of Weyl semimetals. Electric and magnetic multipole analysis provides a deeper understanding of the symmetry-mediated, momentum-dependent spin-splitting, which has strictly non-relativistic origin. GdAlSi is possibly the first candidate material with non-centrosymmetric collinear AFM structure showing such momentum-dependent spin-splitting, as also confirmed by our detailed symmetry analysis. To the best of our knowledge, such co-existence of unconventional AFM order and non-trivial topology is unprecedented and has not been observed before in a single material, rendering GdAlSi a special and promising candidate material. We further propose a device harnessing these features, poised to enable practical and efficient topotronic applications.
\end{abstract}

\maketitle

\section{Introduction}
    
A large number of today's technological advancements are based on extensive research in condensed matter and material physics, which bring about innovations in the field of electronics. The control and manipulation of electronic charge using different devices dominates the functioning of our day-to-day lives and hence it requires a constant push towards the hunt for more power-efficient devices with scalable architecture and simple configurations. Spintronics brings a key player along these lines, which is dictated by the intrinsic spin of electrons. The tunability of this extra degree of freedom along with the charge of electrons have proved to be promising with several proof-of-concept devices, taking us a few steps closer to surpassing semiconductor junctions in mainstream technological applications.

Conventionally, materials with ferromagnetic (FM) order have been the primary choice for spintronic based devices due to their inherent spin polarization \cite{inomata2008highly, hirohata2014future}, a consequence of exchange splitting. But these materials are sensitive to external stray fields when used in devices. Materials with antiferromagnetic (AFM) order can overcome such shortcomings \cite{jungwirth2016antiferromagnetic, baltz2018antiferromagnetic} , but for collinear ordering, there is a distinct absence of any spin polarization due to the mirror symmetric density of states arising out of combined inversion ($\mathcal{P}$) and time-reversal ($\mathcal{T}$) symmetries. This can be overcome via specific materials hosting non-collinear AFM order which lifts the spin degeneracy along with a fully compensated magnetic moment \cite{jungwirth2016antiferromagnetic}. Interestingly, recent years have witnessed yet another class of materials which acquire certain symmetries in the crystal and magnetic structures that can lead to spin-splitting in the momentum space of a collinear AFM system \cite{Libor2020, Naka2019, Kyo-Hoon2019, Hayami2019, Yuan2020, Yuan2021, Smejkal2022, PhysRevX.12.040501, Mazin2022, Yuan, yuan2023uncovering}. RuO$_2$ is the classic example \cite{Libor2020, Shao2021, Smejkal2022, PhysRevX.12.040501} among these. Such momentum-dependent spin-splitting with intimate connection to crystal symmetries can be used as efficient spin-splitters leading to better spintronic devices\cite{Hayami2019,Gonzalez2021,Bai2022,Hayami2022,Karube2022,Bose2022}.

In the past few decades, the discovery of topological materials with ideally dissipationless, symmetry-protected, and robust spin-polarized surface states (SS) have garnered immense interest due to their potential spintronic applications, leading to the broader topic of topotronics. These topological materials are categorized into different classes based on their topological invariants. Weyl semimetals (WSMs) are one such class with either broken $\mathcal{T}$ or $\mathcal{P}$ symmetry, that hosts Weyl Fermions as low-energy quasi-particle excitations. In WSMs, the bulk valence and conduction bands cross each other at discrete points (always occur in pairs) of definite chirality (handedness), known as Weyl points, while having linear dispersion around the node along all three momentum directions. Though separated in the bulk, these pairs of Weyl points give rise to spin-polarized Fermi arcs on the surface of a WSM which originate and terminate at the surface-projection of one such pair of Weyl points of opposite chiralities. These characteristic features lead to the observation of exotic phenomena such as the chiral anomaly, unconventional anomalous Hall effect, and unusual magnetoresistance. Discovery of Weyl Fermions has been reported first in non-magnetic systems such as TaAs \cite{lv2015observation, PhysRevX.5.011029, xu2015discovery} and 
transition-metal binary/ternary chalcogenides with tilted (type-II) Weyl cones \cite{soluyanov2015type, deng2016experimental, wu2016observation, PhysRevB.93.201101,PhysRevB.95.241108} followed by magnetic systems such as Mn$_3$Sn \cite{kuroda2017evidence}, Co$_3$Sn$_2$S$_2$ \cite{wang2018large}, and Co$_2$MnGa \cite{belopolski2019discovery}. 
Recently, unconventional WSMs have been reported in a few non-centrosymmetric magnetic systems, RAlSi/Ge (R: rare earth) \cite{xu2017discovery, PhysRevLett.124.017202, gaudet2021weyl, li2023, sanchez2020observation}. These systems exhibit diverse electronic properties, including non-trivial topology and strong electron correlations \cite{xu2017discovery, sanchez2020observation}. As such, they provide a great platform to investigate the interplay between Weyl Fermions and exotic magnetic order, as both the broken $\mathcal{P}$ and $\mathcal{T}$ symmetries play a role in mediating Weyl fermionic states. In contrast to previous reports on RAlSi/Ge where the light rare earth elements have smaller magnetic moments ($m$), GdAlSi with the largest moment ($m\sim7.95 \mu_B$) is the ideal candidate to study such an interplay.

In this article, we present a combined first-principles calculation, angle-resolved photoemission spectroscopy (ARPES) and magneto-transport studies of the emergent topological phases in non-centrosymmetric unconventional antiferromagnetic WSM GdAlSi. The purpose of this study is two-fold. Firstly, we predict the presence of momentum-dependent spin-splitting with a strong non-relativistic origin in a non-centrosymmetric collinear AFM phase as confirmed from symmetry analysis and electric and magnetic multipole studies \cite{Naka2019,Hayami2019}. This effect is sometimes nomenclatured as `altermagnetism' \cite{Smejkal2022, PhysRevX.12.040501, Mazin2022, Libor2020, Naka2019, Kyo-Hoon2019, Hayami2019, Yuan2020, Yuan2021, Yuan, yuan2023uncovering} which has never been observed in a non-centrosymmetric system earlier.
Secondly, we predict the presence of an unconventional WSM phase in GdAlSi using \textit{ab-initio} calculations and verify the same using the ARPES measurements. 
We also analysed the effects of such momentum-dependent splitting on non-trivial band topology. Overall, GdAlSi is the first exclusive candidate material which co-hosts both altermagnetism and topological non-trivial features arising out of a unique interplay between magnetic order, topology and electronic correlations. Finally, we propose a device architecture which can leverage such unique interplay for practical spintronic/topotronic characteristics. Our systematic study not only contributes to a better understanding of non-trivial topology of GdAlSi but also sheds new light on the intimate relation between magnetic order and topology.\\

\section{Methods}
    
\subsection{Single crystal growth}
High-quality single crystals of GdAlSi were grown using the self-flux method in regular alumina crucibles with an elemental ratio of Gd:Al:Si = 1:15:1, each having a purity of 99.99$\%$. The alumina crucibles were sealed in quartz ampules and heated up to 1000$^{\circ}$ C for 36 hours at a rate of 3$^{\circ}$C/hour and were slowly cooled down to 700 $^{\circ}$C at a rate of 0.05$^{\circ}$ C/min for the centrifuge to remove the residual Al flux.

\subsection{Angle-resolved photoemission spectroscopy (ARPES)}
ARPES measurements were conducted at the BL-1 beamline in the Hiroshima Synchrotron Radiation Center (HiSOR), utilizing a hemispherical electron analyzer (VG-SCIENTA, R4000), in the vacuum ultraviolet (VUV) range of 40-65 eV. The base pressure of the ARPES chamber was maintained at 4$\times$ 10$^{-9}$ Pa. 
Energy and angular resolutions for ARPES were set at $\sim$ 40 meV and $\sim$ $\pm$ 0.1$^{\circ}$, respectively. The beam spot size was $\sim$ 10$\times$30 $\mu$m. The temperature during measurements was kept below 30 K.

\subsection{Computational details}
To investigate the electronic structure of GdAlSi in different magnetic configurations, $\textit{ab-initio}$ calculations were performed within the density functional theory (DFT) framework \cite{hohenberg_kohn1964, kohn_sham1965} using the Vienna Ab-initio Simulation Package (VASP) \cite{kresse1996efficient, kresse1996efficiency, kresse1993ab}, based on the projector-augmented wave (PAW) \cite{blochl1994, kresse1999ultrasoft} method.
The Perdew, Burke, and Ernzerhof (PBE) \cite{perdew1996generalized} exchange-correlation potential within the generalized gradient approximation (GGA) was employed.  Calculations are performed both with and without spin-orbit coupling (SOC). A Hubbard U of 7.0 eV was applied on the Gd site using Dudarev's method \cite{dudarev1998} to capture the correlation effects. Brillouin zone (BZ) integration was performed on a 12$\times$12$\times$12 $\Gamma$-centered $k$-mesh using the tetrahedron method with the total energy convergence criteria set to $10^{-8}$ eV. A plane wave energy cutoff of 380 eV was used for all the calculations. The experimental crystal structure was fully optimized until the total forces on individual atoms were less than 0.001 eV/\AA. A tight-binding Hamiltonian was constructed from pre-converged DFT results using the Wannier90 \cite{PhysRevB.56.12847, PhysRevB.65.035109, RevModPhys.84.1419, wannier90v1, wannier90v2, wannier90v3} package. A total of 120 bands were wannierized to calculate the (001)-surface states, with projections on atomic sites including Gd ($s, p, d, f$), Al ($s, p$), and Si ($s, p$) orbitals. The iterative Green's function approach \cite{lee1981_1, lee1981_2, sancho1984, sancho1985} was used to calculate the surface dispersion and Fermi surface (FS) using a semi-infinite slab as implemented in the WannierTools \cite{wu2018wanniertools} package.
Symmetry analysis of the point and space groups were done using the FINDSYM program \cite{Stokes2005, Stokes2013} and the Bilbao Crystallographic Server \cite{Aroyo2006, Aroyo2014, Gallego2012, Perez-Mato2015}.

To compute the atomic-site charge ($\cal T$ even) and magnetic ($\cal T$ odd) multipoles, we decomposed the calculated density matrix $\rho_{lm,l'm'}$ \cite{Cricchio2009, Spaldin2013} into tensor moments. The $\mathcal{P}$ symmetric (even parity) multipoles have contributions from even $l+l'$ terms, while the broken $\mathcal{P}$ symmetric (odd parity) multipoles have only contributions from odd $l+l'$ terms. For the even and odd parity multipoles at the Gd atoms, we, therefore, evaluated $f-f$ and $f-d$ matrix element contributions.

\section{Results and discussion}
     
\subsection{Structural properties} \label{cryst}
\begin{figure*}[t]
    \centering
    \includegraphics[width=1.0\linewidth]{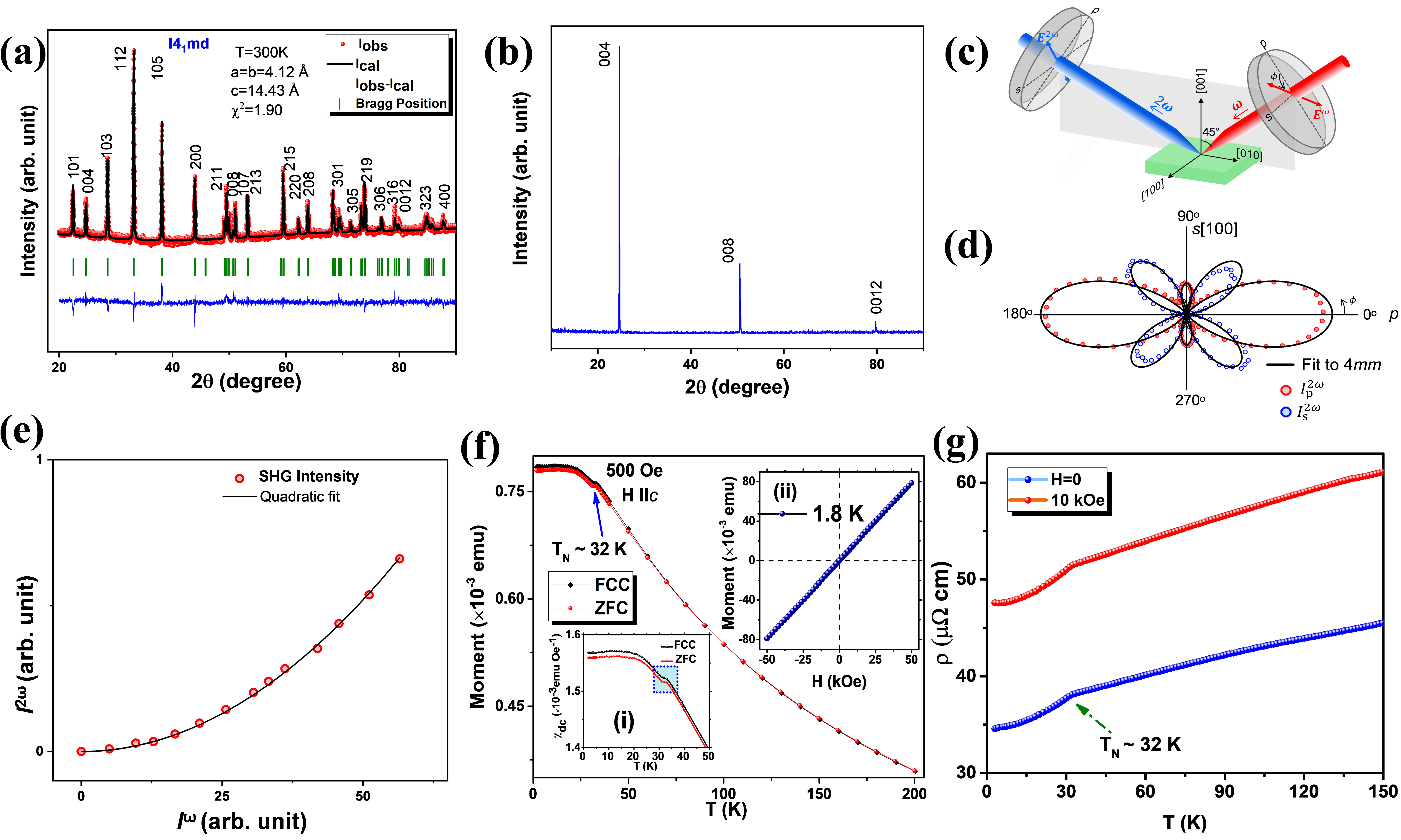}
    \caption{For GdAlSi (a) XRD along with the Rietveld refined data on powder sample. (b) The observed sharp (00$l$) peaks in single crystal XRD pattern imply high quality of the sample. (c) Schematic of the second harmonic generation (SHG) experimental geometry: An 800 nm fundamental wavelength is employed to generate 400 nm second harmonic light in reflection geometry, with the angle of incidence set at 45$^o$.(d) Measured SHG intensity as a function of the incident polarization angle ($\phi$) of the fundamental beam at two fixed analyzer positions (parallel to \textit{p} and \textit{s} polarization of the incident electric field). The experimental data is shown by circles, while the black solid lines depict the theoretical fit based on point group $4mm$.(e) SHG intensity as a function of incident intensity in the normal reflection geometry along with quadratic fitting, establishing broken inversion symmetry in GdAlSi crystals. (f) Magnetization (M) vs. T with H = 500 Oe (H$\parallel$$c$) in zero field cooled (ZFC) and field cooled cooling (FCC) modes. Insets (i) and (ii) show susceptibility ($\chi$) vs. T and M vs. H at 1.8 K revealing an AFM transition around 32 K (see blue dashed square of inset (i)). (g) Resistivity ($\rho$) vs. T at two different fields. A clear kink is observed, confirming ordering temperature (T$_N$=32 K).}
    \label{fig:figs02}
\end{figure*}

Figure \ref{fig:figs02}(a) shows the measured XRD-data along with the Rietveld refinement of GdAlSi, confirming the body-centered tetragonal (BCT) structure with lattice parameters $a=b=4.12$ {\AA} and $c=14.43$ {\AA}. Figure \ref{fig:figs02}(b) shows single crystal XRD with sharp (00$l$) peaks. The structure consists of two Gd, two Al, and two Si atoms in the primitive BCT cell. A more detailed structural properties analysis using Rietveld refinement along with Laue pattern and elemental analysis are shown in Fig. S1 and Table S1 of SM \cite{supplement}.

In general, RAlSi compounds are known for exhibiting Al/Si site disorder, where Al and Si atoms can randomly occupy each other's Wyckoff positions. This disorder results in a centrosymmetric space group $I4_{1}/amd$ ($141$), as opposed to a non-centrosymmetric space group $I4_{1}md$ ($109$) if Al and Si retain their respective Wyckoff positions. To find the actual structure, we performed optical second harmonic generation experiments (see Figure \ref{fig:figs02}(c) for schematic of the SHG experimental geometry) on GdAlSi crystal. SHG is a non-linear optical process that doubles the frequency, where two photons at the fundamental frequency ($\omega$) produce a photon with twice the frequency (2$\omega$) \cite{paperr}. Since only non-centrosymmetric materials are dipolar SHG active, SHG serves as an excellent probe for detecting broken inversion symmetry.  The combined results from XRD refinement and strong SHG signal (see Fig. \ref{fig:figs02}(e)) along with SHG polarimetry fitting \textcolor{black}{(see Sec. S1(C-D) of SM \cite{supplement} for details of SHG polarimetry and fitting) as shown in Figure \ref{fig:figs02}(d)}, confirm that GdAlSi crystallizes in the non-centrosymmetric space group $I4_{1}md$ (point group \( 4mm \)).

\begin{figure*}[!htb]
    \centering
    \includegraphics[width=1.0\linewidth]{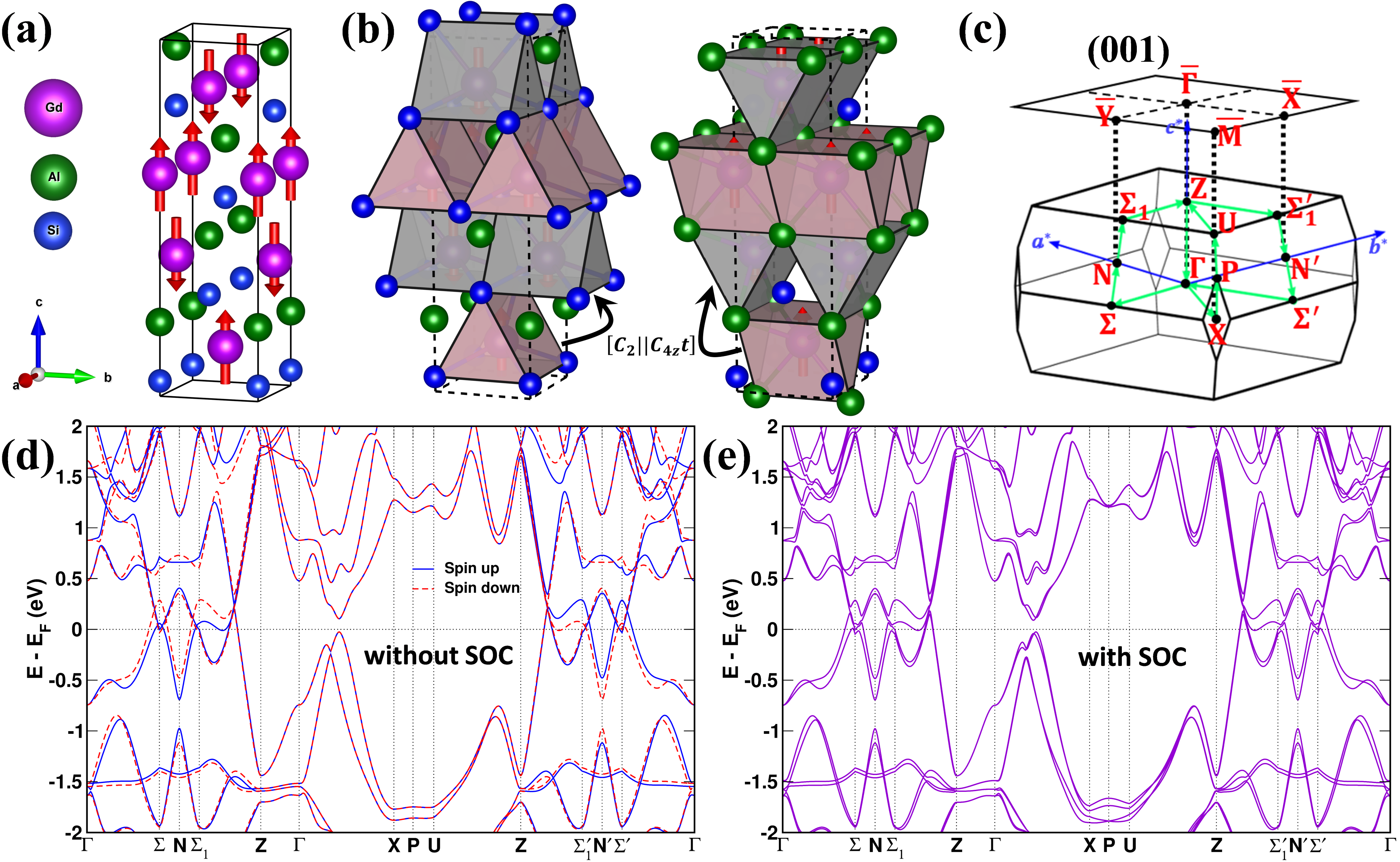}
    \caption{For GdAlSi, (a) Tetragonal crystal structure with lowest energy magnetic ordering (AFM type-I). (b) Inequivalent crystal environment around the two Gd-sites shown by pink and grey prism-like polyhedra around them. Left and right figures show a unique $[C_2||C_{4z}\textbf{t}]$ rotated version of one over the other, highlighting GdAl$_6$, GdSi$_6$ units. (c) Primitive bulk and surface (001) Brillouin zone (BZ).  (d) Spin-polarized bulk band structure without SOC for AFM type-I configuration. Spin up (down) states are shown by solid (dashed) lines. Substantial band splitting is evident along certain k-paths. (e) Bulk band structure with SOC for AFM type-I configuration along the same path as in (d) highlighting the masking of SOC effect by non-relativistic spin-splitting.}
    \label{fig:fig1}
\end{figure*}

\subsection{Magnetization} \label{mag}

Figure \ref{fig:figs02}(f) shows magnetization (M) vs. temperature(T) and field (H) data along with the dc magnetic susceptibility for GdAlSi. It shows a noticeable deviation at $\sim$ 32 K, indicated by a kink that corresponds to the onset of AFM ordering (see the highlighted blue dashed square of inset (i)). \textcolor{black}{The sharpness of the transition may indicate the collinear structure.} This observation is further supported by the analysis of the resistivity data (see \ref{fig:figs02}(g)). Above the N\'{e}el temperature ({\it{T}}$_N$ = 32 K), the resistivity and susceptibility exhibit consistent behavior \cite{bobev2005ternary}. The Curie-Weiss law fit yields an effective moment of 7.95 $\mu_B$, as expected for Gd$^{3+}$, along with a Weiss constant of -103 K. Non-saturating behavior (even up to 50 kOe field) of low temperature M vs. H curve (see inset of Fig. \ref{fig:figs02}(f)) indicates the presence of AFM interactions in the system.

\subsection{\textit{Ab-initio} results} \label{bulk}

In order to investigate the electronic structure of GdAlSi, paramagnetic and various magnetic states including ferromagnetic, ferrimagnetic, antiferromagnetic, and non-collinear helical configurations were simulated to determine the magnetic ground state. Among these, the AFM type-I configuration (interlayer AFM ordering among Gd atoms with moments oriented along the $c$-axis as shown in Fig. \ref{fig:fig1}(a)) is found to be the energetically most favorable configuration (also see Fig. S2 and Table S2 of SM \cite{supplement}). This configuration arises from the alignment and coupling of localized $f$-electrons of Gd along the $c$-axis, leading to spin-polarization of the conduction electrons. 

An important feature of the crystal structure (Fig. \ref{fig:fig1}(b)) is the inequivalent environments of Al and Si atoms around the Gd sublattices forming a prism-like network. In particular, both GdAl$_6$ and GdSi$_6$ units around the two Gd sublattices are related to each other by a $C_{4z}$ rotational symmetry (axis of rotation coinciding with the $c$-axis of the tetragonal cell). More interestingly, the GdAl$_6$ and GdSi$_6$ units around the same Gd sublattice are also $90^\circ$ rotated with respect to each other, as depicted by pink and grey prism-like polyhedra in Fig. \ref{fig:fig1}(b). These inequivalent environments play a crucial role in driving the non-relativistic spin-splitting (NRSS) as well as its manipulation in GdAlSi.

Figure \ref{fig:fig1}(d) depicts the spin-polarized electronic bulk band structure of GdAlSi without spin-orbit coupling (SOC) using theoretically optimized lattice parameters ($a=b=4.14$ \AA~ and $c=14.50$ \AA). The emergence of hole pockets along $\Gamma-\Sigma-\text{N}-\Sigma_1-\text{Z}$ and $\Gamma-\Sigma^{\prime}-\text{N}'-\Sigma_1^{\prime}-\text{Z}$ path indicates a semi-metallic nature along with a large spin-splitting ($\sim$230 meV at certain momenta) which is of particular interest. Notably, such splitting is highly path-dependent (e.g. the bands along $\text{Z}-\Gamma-\text{X}-\text{P}-\text{U}-\text{Z}$ are completely spin-degenerate) and the spin-splitting flips when going from $\Gamma-\Sigma-\text{N}-\Sigma_1-\text{Z}$ direction to $\Gamma-\Sigma^{\prime}-\text{N}'-\Sigma_1^{\prime}-\text{Z}$ direction in the BZ. These two directions are at 90$^{\circ}$ to each other as evident from Fig. \ref{fig:fig1}(d). Presence of such \textit{alternating} spin-polarization in a collinear AFM system, while absence of the same along a specific path in the BZ gives us the first hint towards a possibility of the so-called \textit{altermagnetism} \cite{Smejkal2022, PhysRevX.12.040501, Mazin2022} in GdAlSi. This implies a spin-split bulk Fermi surface with the spin-up and the spin-down FS being rotated by 90$^{\circ}$ with respect to each other as a consequence of crystalline symmetries.  A projection of the bulk FS on the $k_z=0$ plane is shown in Fig. S3 of SM \cite{supplement} which clearly illustrates this. Interestingly, even in the absence of strong correlations arising out of Gd-$f$ electrons, GdAlSi retains the  \textit{altermagnetic} band structure in AFM-I configuration (see Figs. S4(a,b) of SM \cite{supplement}). In the paramagnetic phase of GdAlSi without considering SOC, the electronic bands are doubly degenerate throughout the BZ due to the presence of $\mathcal{PT}$ symmetry in the momentum space (bands are always $\mathcal{P}$-symmetric in the absence of SOC, see Figs. S4(c,d) and associated discussions in section S2 of SM \cite{supplement}). This implies that the alternating spin-splitting is a consequence of spin-space and real-space crystalline symmetries resulting in the breaking of $\mathcal{T}$ symmetry due to AFM-I order.\\ 

\begin{figure*}[!htb]
    \centering
    \includegraphics[width=1.0\linewidth]{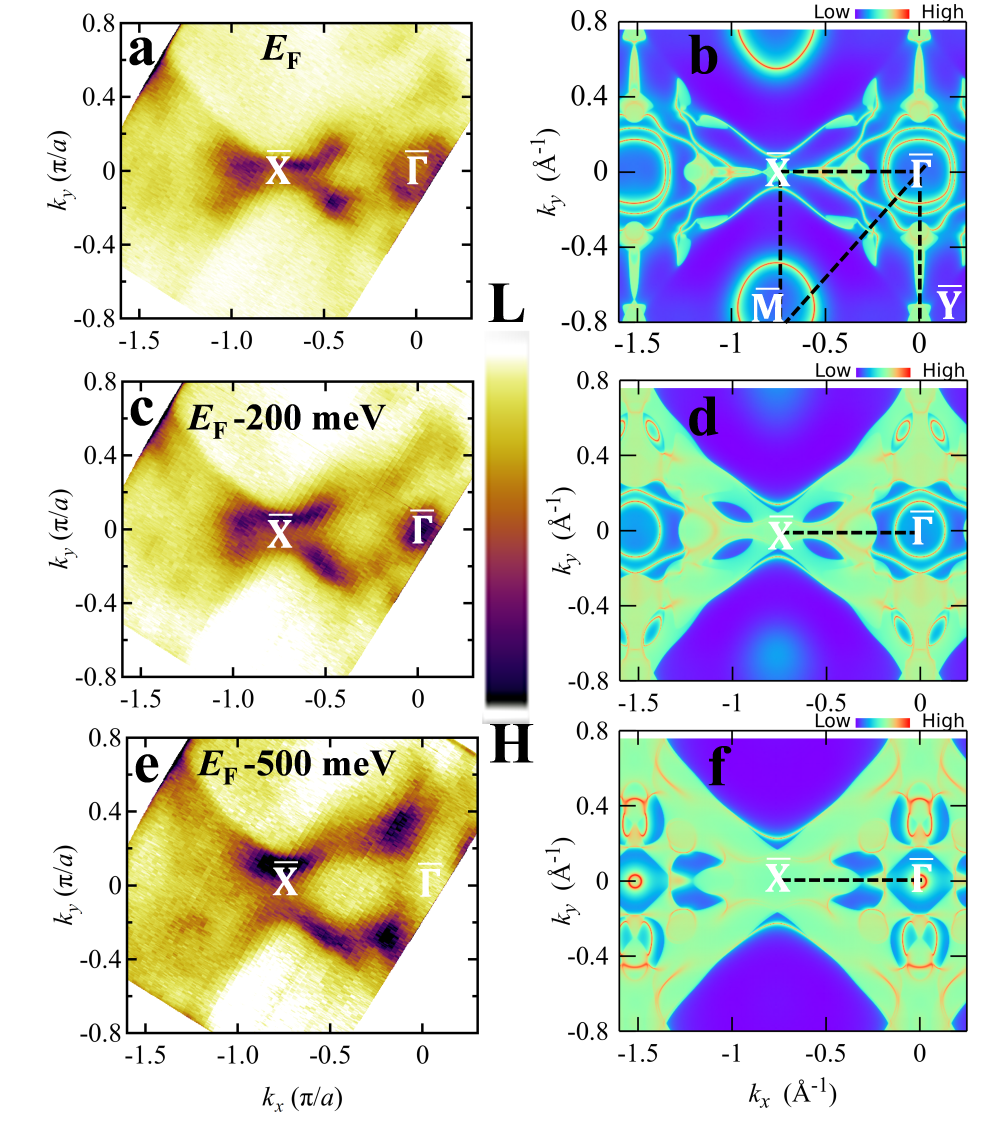}
    \caption{For GdAlSi, distribution of 32 Weyl nodes within the bulk BZ observed from (a) side and (b) top ((001) direction) view in the presence of SOC. Weyl nodes with +1 (-1) chirality are shown by red (blue) dots and the $k_z=0$ plane is highlighted in green. Dashed boxes highlight the type-I and II WPs. 3D bulk Fermi surface in the presence of SOC as viewed from (c) side and (d) top. Electron and hole pockets are shown in red and blue respectively. Bulk electronic dispersions in the vicinity of (e,f) type-I and (g,h) type-II WPs. The bands forming the Weyl cones have been highlighted in magenta.}
    \label{FS}
\end{figure*}

\textcolor{black}{To understand the symmetry-mediated origin of altermagnetism in GdAlSi better, we begin with the point group of GdAlSi (derived from the parent space group $I4_{1}md$) which is $4mm\;(C_{4v})$. Altermagnetism requires a two-coset decomposition of the parent point group using an index-2 (halving) subgroup containing the same-spin and an opposite-spin sublattice transformation generating the non-trivial coset from the subgroup, such that the coset does not contain the spatial inversion element $\mathcal{P}$ \cite{Smejkal2022, PhysRevX.12.040501}. The point group $C_{4v}$ has two halving subgroups, namely $4\;(C_{4})$ and $mm2\;(C_{2v})$. From the crystal structure shown in Figs. \ref{fig:fig1}(a) and \ref{fig:fig1}(b), it is evident that the opposite-spin sublattices are related to each other by a $C_{4z}$ rotation. Hence, $C_{4z}$ should generate the non-trivial coset containing all the possible opposite-spin sublattice transformations. This makes $C_{2v}$ as the halving subgroup. Since $C_{4v}$ is a non-centrosymmetric point group, $C_{2v}$ and its corresponding coset trivially does not contain $\mathcal{P}$. Within the spin point group formalism \cite{Litvin1977}, the non-trivial spin group for GdAlSi takes the following form:
\begin{equation}
    R_s = [\,\mathbb{I}\,||\,C_{2v}\,]\cup[\,C_2\,||\,C_{4z}\,]\,[\,\mathbb{I}\,||\,C_{2v}\,]
    \label{eq1}
\end{equation}
where, $\mathbb{I}$ is the spin-space identity transformation, $C_2$ acts only on the spin-space (leading to spin-space inversion) and $C_{4z}$ acts only on the real-space (position and momentum space). It must be noted that while $\mathbb{I}$, $C_2$ and $C_{4z}$ are group elements (transformations), $C_{2v}$ and $C_{4v}$ are point groups containing real-space transformations. Among a total of 8 transformation elements, there are four mirror transformations present in $C_{4v}$, namely: $\mathcal{M}_{x}, \mathcal{M}_{y}, \mathcal{M}_{xy}, \mathcal{M}_{x\overline{y}}$. After the two-coset decomposition, $\mathcal{M}_{x}$ and $\mathcal{M}_{y}$ mirrors remain in the subgroup $C_{2v}$ whereas $\mathcal{M}_{xy}$ and $\mathcal{M}_{x\overline{y}}$ go into the non-trivial coset. The coset also contains the 4-fold rotations $C_{4z}$ and $C_{4z}^3$. It is evident from Eq.\eqref{eq1} that the effect of different spin sublattices will be generated by the coset transformations. This has interesting consequence on the electronic band structure of GdAlSi. If we denote the energy eigenvalues at a generic momentum $\Vec{k}=(k_x,k_y,k_z)$ as $\varepsilon_n(s,\Vec{k})$ indexed by spin $s$ and $R$ as a coset transformation element, then the following relation will hold in general:
\begin{equation}
    [\,C_2\,||\,R\,]\,\varepsilon_n(s,\Vec{k}) = \varepsilon_n(-s,\Vec{k}') = \varepsilon_n(s,\Vec{k})
    \label{eq2}
\end{equation}
where, $\Vec{k}'=R\,\Vec{k}$ and the last equality holds since $[\,C_2\,||\,R\,]$ is a symmetry of the system. Considering the coset transformations, it follows:
\begin{equation}
    \Vec{k}' =
    \begin{cases}
        (-k_y,-k_x,k_z) & R = \mathcal{M}_{xy} \\
        (k_y,k_x,k_z)   & R = \mathcal{M}_{x\overline{y}} \\
        (k_y,-k_x,k_z)  & R = C_{4z} \\
        (-k_y,k_x,k_z)  & R = C_{4z}^3
    \end{cases}
    \label{eq3}
\end{equation}
From Eq.\eqref{eq2} and Eq.\eqref{eq3}, it can be concluded that energy bands along the $\text{Z}-\Gamma-\text{X}-\text{P}-\text{U}-\text{Z}$ direction will be spin-degenerate as they lie on the $\mathcal{M}_{x\overline{y}}$ mirror plane ($k_x=k_y$ on the plane). The other mirror plane $\mathcal{M}_{xy}$ ($k_x=-k_y$ on the plane) will also be spin-degenerate (not shown here). The spin-degeneracy of energy bands along the $\Gamma-\text{Z}$ direction is additionally protected by $C_{4z}$ and $C_{4z}^3$ symmetries. But for other generic momenta, such as those along $\Gamma-\Sigma-\text{N}-\Sigma_1-\text{Z}$ and $\Gamma-\Sigma^{\prime}-\text{N}'-\Sigma_1^{\prime}-\text{Z}$ paths, the spin-degeneracy will be lifted. It is to be noted that the coset does not contain real-space identity transformation which implies absence of $[\,\mathcal{T}\,||\,\mathcal{T}\,]$, i.e., broken $\mathcal{T}$ symmetry. This further facilitates the lifting of spin-degeneracy at/along these other generic momenta. Thus, GdAlSi will show $d$-wave altermagnetism with the spin point group \cite{Litvin1977} given by $^{2}4^{1}m^{2}m$ and characteristic spin group integer 2 \cite{Smejkal2022, PhysRevX.12.040501}. Here, $^{2}4$ and $^{2}m$ arise from the coset whereas $^{1}m$ comes from the halving subgroup. Considering the full spin group \cite{Litvin1977, Litvin1973, Litvin1974}, the 4-fold rotation and the second mirror plane in $^{2}4^{1}m^{2}m$ become the screw rotation $4_{1}$ and the $d$ (diamond) glide mirror plane of the parent space group $I4_{1}md$ respectively. The full transformation between opposite-spin sublattices is then generated by $C_{4z}\textbf{t}$ yielding the spin group element $[\,C_2\,||\,C_{4z}\textbf{t}\,]$, where the 4-fold rotation $C_{4z}$ is followed by a non-symmorphic translation $\textbf{t}$ in the real-space as shown in Fig. \ref{fig:fig1}(b).}
\textcolor{black}{The mentioned group symmetries also dictate the presence or absence of multipoles in a system. We further discuss how these multipoles may affect and possibly manipulate the \textit{altermagnetic} spin-splitting later in section \ref{nrss}.}

\begin{figure*}[!htb]
    \includegraphics[width=1.0\linewidth]{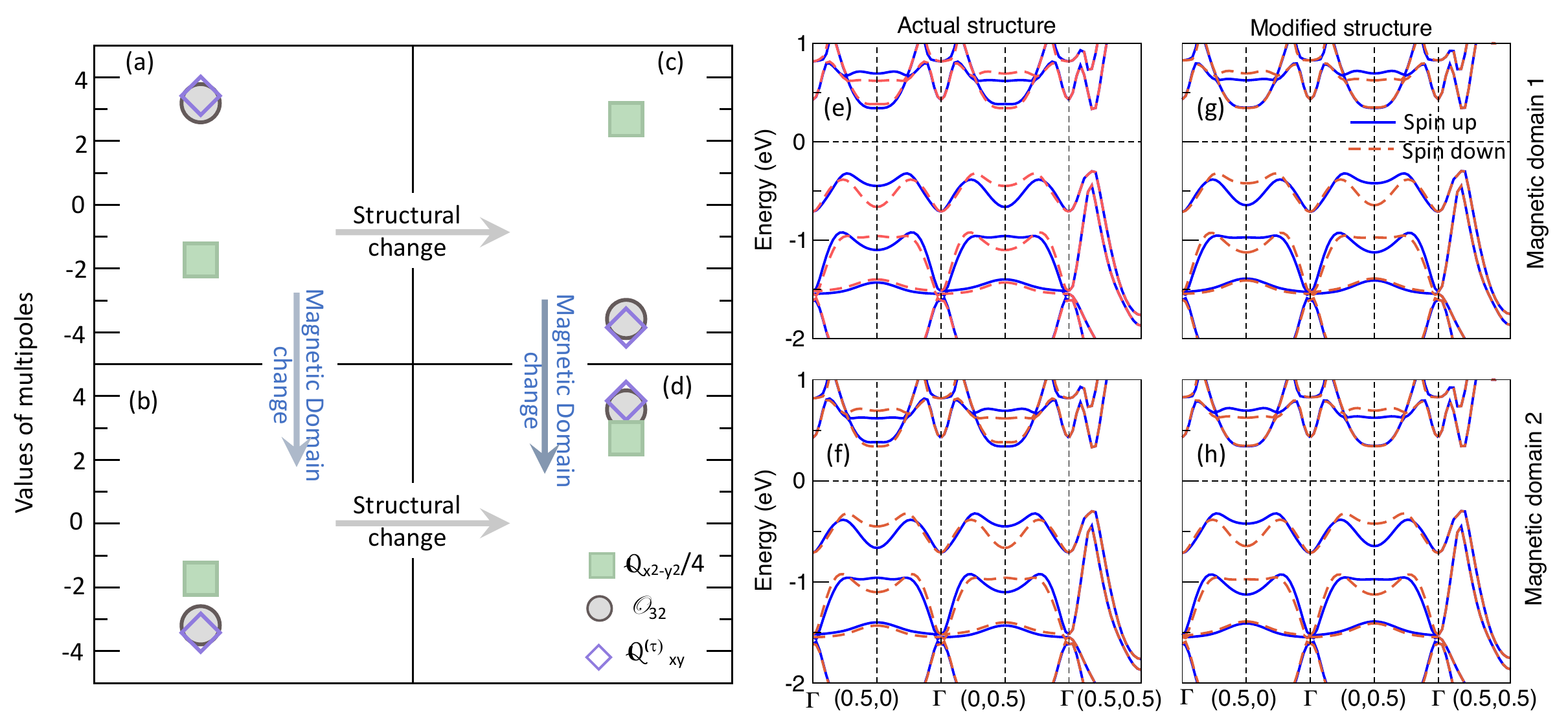}
    \caption{Manipulation of the magnetic octupoles and the NRSS in GdAlSi in absence of spin-orbit coupling (SOC). Computed charge quadrupole moment ${\cal Q}_{x^2-y^2}$ (in units of 0.25 $\times 10^{-4}$ electronic charge) and the magnetic octupole components (in $10^{-4}$ $\mu_B$) ${\cal O}_{32}$ and ${\cal Q}^{(\tau)}_{xy}$ in absence of SOC for (a) the actual crystal structure of GdAlSi with the magnetic configuration shown in Fig. \ref{fig:fig1}, (b) the actual crystal structure but with opposite magnetic arrangement, (c) hypothetical modified structure (with Al and Si positions interchanged) with the same magnetic arrangement as in Fig. \ref{fig:fig1}, and (d) the same hypothetical modified structure with the opposite magnetic arrangement. (e-h) Corresponding spin-polarized band structure (solid and dashed lines indicating the up and down spin-polarized bands respectively) at $k_z=0$ plane.
    }
    \label{fig1}
\end{figure*}

With the inclusion of SOC, the spin group of GdAlSi is reduced to the magnetic space group $I4_{1}'m'd$ with the magnetic point group $4'm'm$ when considering the magnetic moments on Gd atoms to be oriented along the crystalline $c$-axis. In this case, although the electronic bands are slightly perturbed, the overall feature of the bands remain similar with only minor splittings of about $\sim$ 70 meV at certain momenta \textcolor{black}{(see Fig. \ref{fig:fig1}(e))} which is relatively smaller than the non-relativistic spin-splitting ($\sim$ 230 meV). \textcolor{black}{Most notably, the spin-degeneracy along $\Gamma-\text{X}-\text{P}-\text{U}-\text{Z}$ direction is lifted, whereas this effect is masked by the non-relativistic spin-splitting in other directions as shown in Fig. \ref{fig:fig1}(e). 
} 
The paramagnetic phase of GdAlSi in the presence of SOC also shows such small splitting. This minor lifting of spin-degeneracy due to SOC is a consequence of the non-centrosymmetric crystal structure of GdAlSi, i.e., lack of $\mathcal{P}$ symmetry (also see Figs. S4(c-f) and associated discussions in section S2 of SM \cite{supplement}). \textcolor{black}{Interestingly,} the $z$-component of spin \textcolor{black}{projected onto the relativistic band structure in Fig. \ref{fig:fig1}(e)} still shows the same \textit{alternating} feature as seen in non-relativistic bands. This can be attributed to Gd$^{3+}$ being an effective $s$-state ion (Gd$^{3+}: 4f^{7}$, $L=0$) which implies a dominance of non-relativistic effects on the electronic properties of GdAlSi in spite of Gd$^{3+}$ being a heavy ion. This also implies the $z$-component of spin to be a conserved quantity ($m_s$ is a good quantum number) and it is further protected by $C_{\infty}$ symmetry of the spin-only groups \cite{Kitz1965, Litvin1974} due to the collinear magnetic arrangement \cite{Smejkal2022, PhysRevX.12.040501}. 
Due to SOC, avoided crossings appear near the Fermi level ($E_F$) along the $\Gamma-\Sigma-\text{N}-\Sigma_1-\text{Z}$ and $\Gamma-\Sigma^{\prime}-\text{N}'-\Sigma_1^{\prime}-\text{Z}$ paths leading to high intrinsic total Berry curvature. 
\textcolor{black}{The spin-projected relativistic bands and the total Berry curvature are shown in Figs. S5(a) and S5(b) respectively along with associated discussions in section S2 of SM \cite{supplement}.}

Further, a search of topological nodal points in the bulk BZ reveals a total of 16 pairs of Weyl points (WPs) with opposite chirality  as shown in Figs. \ref{FS}(a,b). Out of these 16 pairs, 8 pairs are  type-I while the other 8 pairs are type-II WPs. All the type-I WPs lie on the $k_z=0$ plane whereas the type-II WPs are distributed on four planes at $k_z=\pm0.29$~\AA$^{-1}$ and $k_z=\pm0.36$~\AA$^{-1}$. The 3D bulk Fermi surface is shown in Figs. \ref{FS}(c,d). The type-I WPs can further be divided into 4 pairs lying at an energy $\sim 24$ meV above $E_F$ (referred to as type-IA) and the other 4 pairs lying at an energy $\sim 79$ meV above $E_F$ (referred to as type-IB). Similarly, the type-II WPs can further be divided into 4 pairs lying at an energy $\sim 58$ meV above $E_F$ (referred to as type-IIA) and the other 4 pairs lying at an energy $\sim 65$ meV above $E_F$ (referred to as type-IIB). The dispersion around the different types of WPs are shown in Figs. \ref{FS}(e-h). More details about the WP coordinates can be found in Table S3 of SM \cite{supplement}. The associated topologically non-trivial Fermi arcs are discussed in more details in Sec. \ref{surface}.

\subsection{NRSS and the role of magnetic octupoles} \label{nrss}

For a deeper understanding of the origin of the NRSS in GdAlSi, and to reveal its characteristics and possible manipulation, we explicitly carry out the multipole analysis.     
In the absence of any net magnetic dipole moment, the spin-splitting has been proposed to mainly arise from higher-order magnetic multipoles \cite{Hayami2019,HayamiPRB2020,HayamiPRB2021,Bhowal2024,Hayami2022}, among which third-rank magnetic octupole, describing the anisotropy in the magnetization density, is a notable candidate. Unlike the centrosymmetric {\it altermagnets} \cite{Smejkal2022, PhysRevX.12.040501, Mazin2022}, GdAlSi breaks $\mathcal{P}$ symmetry, which, in turn, also allows for ferrotype ordering of odd-parity, second-rank magnetoelectric (ME) multipole component ${\cal Q}_{x^2-y^2}$. This leaves the origin of the NRSS in GdAlSi an open question, which we investigated here for the first time. In fact, the NRSS feature in RAlSi family, as a whole, has never been looked upon and hence might require a revisit for all those systems which are reported in the literature so far.

The computed multipoles in GdAlSi confirm the presence of a ferrotype odd-parity ME multipole component ${\cal Q}^{\rm ME}_{x^2-y^2}$ as well as ferrotype even-parity magnetic octupole components \cite{note1} ${\cal O}_{32}$ and ${\cal Q}^{(\tau)}_{xy}$ (the latter also known as toroidal quadrupole moment), consistent with the $I4_1'm'd$ magnetic symmetry. Interestingly, however, we find that the ferrotype ${\cal Q}^{\rm ME}_{x^2-y^2}$ has a non-zero value only in the presence of SOC, while the even-parity magnetic octupoles ${\cal O}_{32}$ and ${\cal Q}^{(\tau)}_{xy}$ exist even in the absence of SOC (see Fig. \ref{fig1}a). This excludes the former from being the origin of the NRSS, hinting magnetic octupoles to be the lowest-order ferroic magnetic multipole in GdAlSi in the absence of SOC. 

To verify if the magnetic octupoles can describe the NRSS in GdAlSi, we analyze the corresponding reciprocal space representation. The $k$-space representation \cite{Watanabe2018,BhowalSpaldin2021,BhowalPRL2022} of the existing magnetic octupole components viz., $(k_x^2-k_y^2)m_z$ with $m_z$ being the collinear magnetic moment along $\hat z$, provides the following insight into the NRSS. First of all, the representation indicates the occurrence of the spin-splitting in the momentum space with non-zero $k_x$ and $k_y$ (see Fig. \ref{fig1}(e)), provided $k_x \ne k_y$ (see the absence of NRSS along $\Gamma \rightarrow (0.5,0.5)$ in Fig. \ref{fig1}(e)). Secondly, the spin-splitting is symmetric in $\vec k$ (see the band dispersion along $\Gamma \rightarrow (0.5,0) \rightarrow \Gamma$ in Fig. \ref{fig1}(e)) as the representation is even in $k^2$. We note that this is in contrast to the spin-orbit coupling driven Rashba-like spin-splitting which is antisymmetric with respect to $\vec k$. Finally, under the $C_{4z}$ rotation of the momentum direction from $k_x$ to $k_y$, the spin-splitting reverses (band dispersion along $\Gamma \rightarrow (0.5,0.0)$ vs. $\Gamma \rightarrow (0.0,0.5)$ in Fig. \ref{fig1}(e)). Thus, our computed band structure confirms all the characteristics mentioned above, predicted from the $k$-space representation of the existing magnetic octupole moments, further providing a convenient description of the NRSS in terms of magnetic octupoles.

It is crucial to highlight that although both magnetic octupoles and the spin-splitting are dictated by the underlying crystal and magnetic symmetries of the material, correlating the NRSS to magnetic octupoles has the advantage that it provides a platform for controlling the NRSS by manipulating the magnetic octupoles \cite{Bhowal2024}. To illustrate this, we first consider the opposite magnetic domain by flipping the direction of the spins without altering the magnetic symmetry. This results in a sign-switching of the magnetic octupole as depicted in Fig. \ref{fig1}(b). Consequently, the spin-splitting also reverses (see Fig. \ref{fig1}(f)). Since the inequivalent Gd$A_6$ ($A=$ Si, Al) environment is responsible for broken ${\cal T}$ symmetry, we further artificially modify the crystal structure by interchanging the atomic positions of Al and Si to see the impact of the surrounding non-magnetic ions on the magnetic octupoles and hence the NRSS. This interchanges the inequivalent GdSi$_6$ and GdAl$_6$ networks, which, in turn, results in a sign change of the charge quadrupole moment ${\cal Q}_{x^2-y^2}$ (see Fig. \ref{fig1}(c)) at the Gd ions. Interestingly, in this case, even without any change in the magnetic arrangement, we find that the magnetic octupoles also change sign (see Fig. \ref{fig1}(c)) and, accordingly, the NRSS also reverses (see Fig. \ref{fig1}(g)), emphasizing the role of the surrounding non-magnetic environment on the NRSS. For this hypothetical modified structure, a change in the magnetic domain further switches the sign of the magnetic octupoles without affecting the charge quadrupole ${\cal Q}_{x^2-y^2}$ (see Fig. \ref{fig1}(d)). As a result, the spin-splitting reverses again (see Fig. \ref{fig1}(h)), and we get a similar band structure as in Fig. \ref{fig1}(e). 

\begin{figure}[!t]
    \centering
    \includegraphics[width=1.0\linewidth]{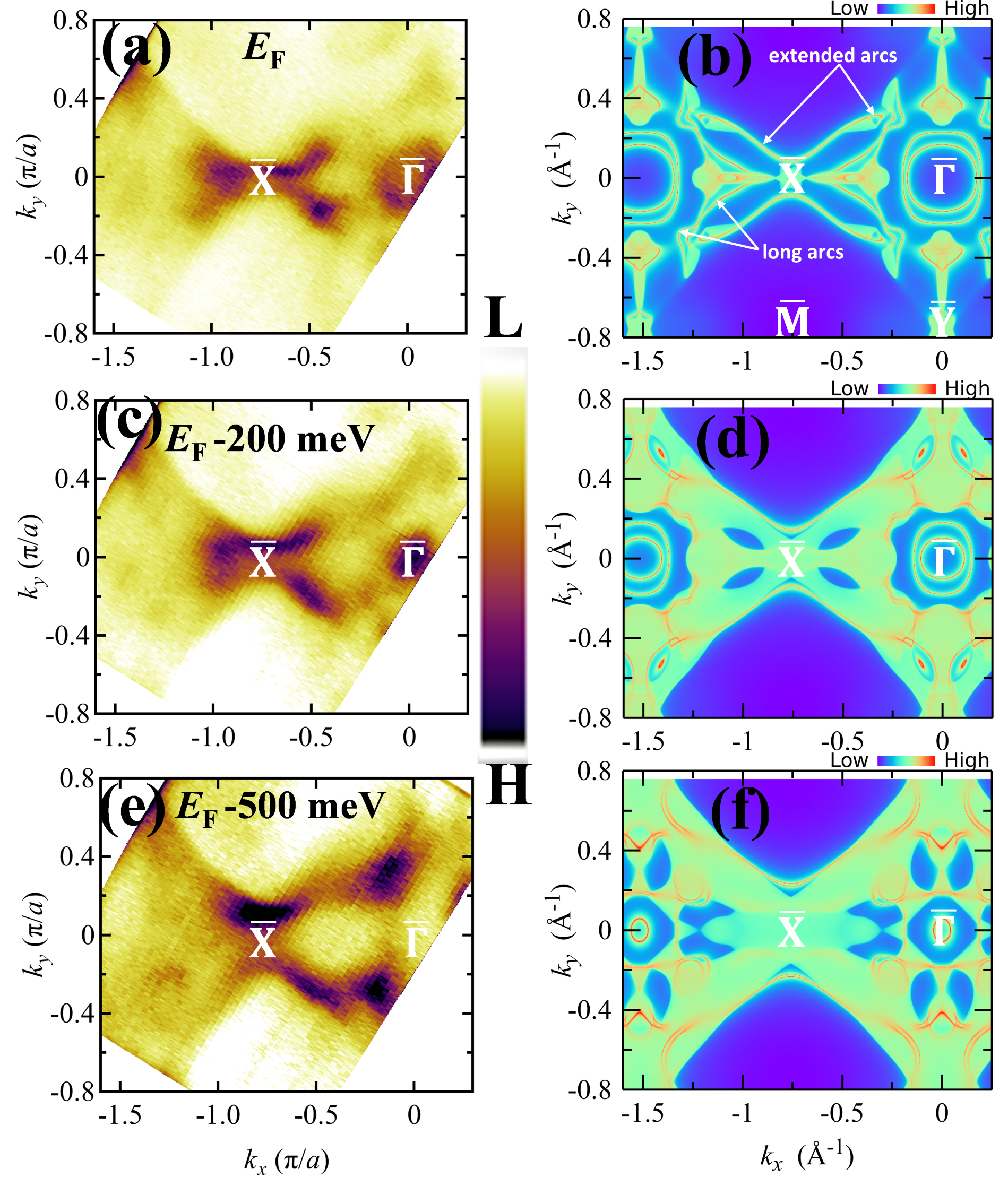}
    \caption[ARPES data (Fermi surfaces) with $h\nu$=50 eV ]{For the (001) surface of GdAlSi,  unsymmetrized measured Fermi surface (FS) and constant binding energy contours at (a) $E_F$ (c) $E_F-200$ and (e) $E_F-500$ meV with $h\nu$ = 50 eV at 24 K. (b,d,f) The corresponding simulated FS and isoenergy contours. }
    \label{fig:fig6}
\end{figure}

\subsection{Surface dispersion and Fermi surface} \label{surface}

\begin{figure}[t]
    \centering
    \includegraphics[width=1.0\linewidth]{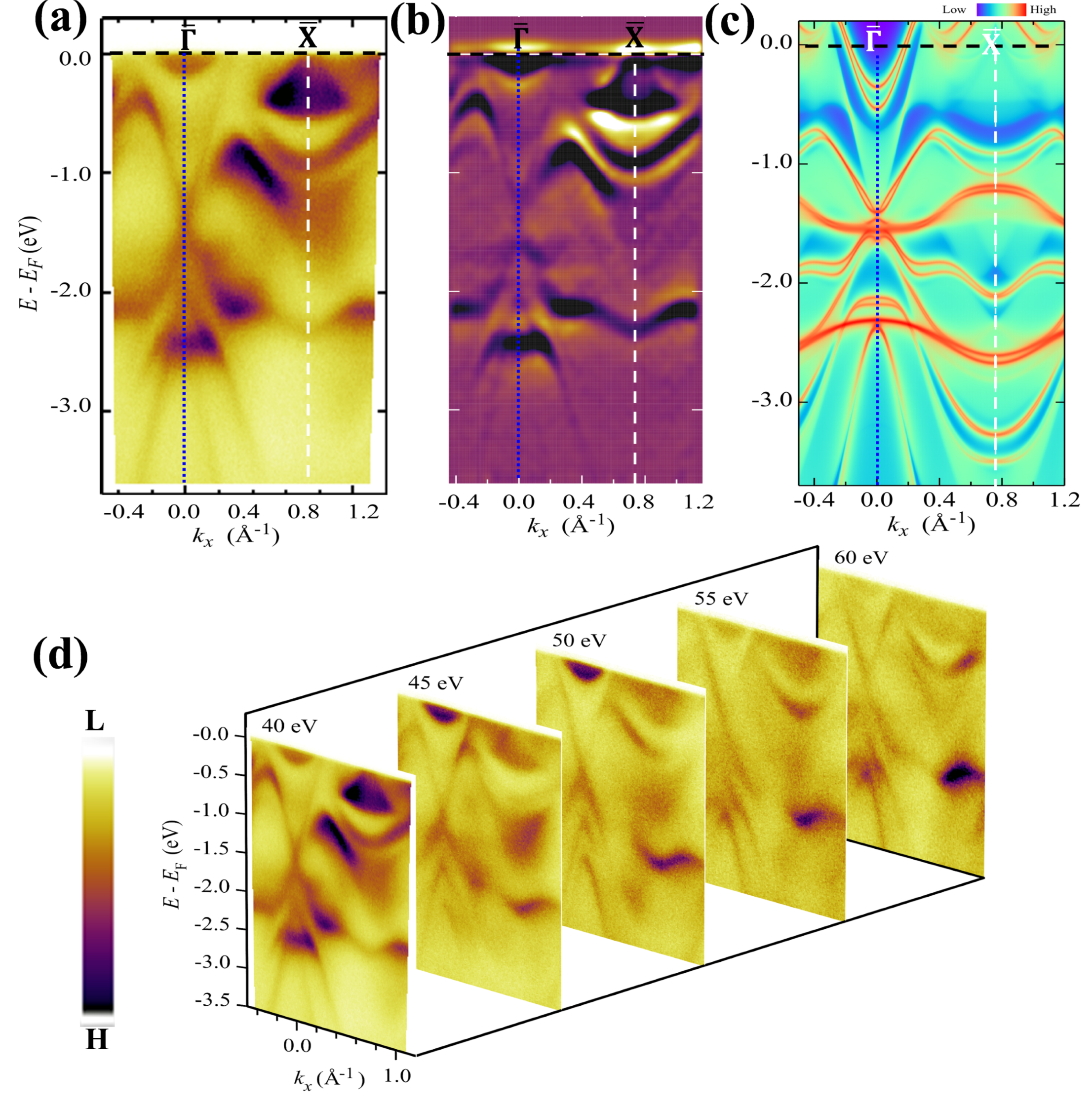}
    \caption[Photon energy dependent band dispersion]{For the (001) surface of GdAlSi, (a) measured ARPES band dispersion at $h\nu=40$ eV (b) a second derivative image of Fig. (a), and (c) calculated surface band dispersion. The blue vertical dotted line and white vertical dashed line locate the $\overline{\Gamma}$ and  $\overline{\text{X}}$-point respectively, while the black horizontal dashed line marks the $E_F$. (d) Photon energy dependent (40-60 eV) band dispersion along $\overline{\Gamma}-\overline{\text{X}}$ line.}
    \label{fig:fig3}
\end{figure}

\textcolor{black}{In GdAlSi, due to bulk-boundary correspondence, the projected WPs on (001) cleaved surface (with respect to conventional cell shown in Fig. \ref{fig:fig1}(a)) are expected to show open Fermi arcs.}
Figures \ref{fig:fig6}(b) and \ref{fig:fig3}(c) show the calculated $k_x-k_y$ 2D Fermi surface (FS) and band dispersion along $\overline{\Gamma} \rightarrow \overline{\text{X}}$ surface high-symmetry line respectively in the presence of SOC. This is an Al-terminated (001) surface, obtained by cleaving the Gd$-$Al bond. The FS is diamond-shaped with an apparent 2-fold symmetry and shows several prominent features.  Two concentric closed contours can be seen surrounding the $\overline{\Gamma}$-point which are electron pockets originating exclusively from the surface. \textcolor{black}{Four bulk-like Fermi pockets originating from bulk type-II Weyl cones can be seen along the $\overline{\Gamma}\rightarrow\overline{\text{X}}$ and $\overline{\Gamma}\rightarrow\overline{\text{Y}}$ directions. These are symmetric due to the presence of $C_{4z}$ and mirror symmetries in the bulk. There are also four butterfly-like bulk-like states symmetrically present about the $\overline{\Gamma}-\overline{\text{M}}$ line. Each of these butterfly-like states contain two pairs of projected type-I(A,B) bulk WPs, with each `wing' of the butterfly having one pair of type-I(A,B) WPs of the same chirality (also see Fig. S6 of SM \cite{supplement}).}
\textcolor{black}{The prominent Fermi arcs seen throughout the surface BZ arise from these type-I WPs. A type-IA WP and a type-IB WP of the same chirality is connected to its chiral partner either via long intra-BZ Fermi arcs or extended inter-BZ Fermi arcs. The long Fermi arcs connect type-I WPs within the same BZ passing through the bulk-like Fermi pockets of the type-II Weyl cones. The extended Fermi arcs, on the other hand, connect type-I WPs of adjacent/neighbouring BZ and have a fishtail-like feature along the $\overline{\Gamma}\rightarrow\overline{\text{X}}$ direction. The surface projections of the type-II WPs are eclipsed by bulk-like Fermi pockets and hence any Fermi arcs originating from type-II WPs are masked and difficult to discern.}  
The fishtail-like arcs along the $\overline{\Gamma}\rightarrow\overline{\text{X}}$ direction persists even at deeper energies (below E$_F$) becoming wider and move away from $\overline{\text{X}}$, while the central surface electron pockets slowly diminish as shown in Fig. \ref{fig:fig6}(d,f). Other surface features are masked due to the emergence of surface-projected trivial bulk pockets at deeper energies. The computed FS indicates the existence of topological Fermi arcs in all four quadrants of the square surface BZ, displaying a distinct asymmetry along the $\overline{\Gamma}\rightarrow\overline{\text{X}}$ and $\overline{\Gamma}\rightarrow\overline{\text{Y}}$ lines. 
\textcolor{black}{The calculated surface $E-k$ dispersion along $\overline{\Gamma}\rightarrow\overline{\text{X}}$ path (see Fig. \ref{fig:fig3}c) also shows interesting features. Two electron pockets are clearly visible at $\overline{\Gamma}$-point near $E_F$ originating exclusively from the surface. There is a pronounced crossing at/around -1.5 eV at the edge of the bulk-energy gap and two downward parabolic band dispersion at deeper energy ($\sim$ -2.0 eV, and -2.5 eV respectively) can be seen at $\overline{\Gamma}$. An upward parabolic-like band dispersion originating from the surface states at/around -1 eV and a strong surface state at/around -2.5 eV at $\overline{\text{X}}$ is also visible.}

Along with the special (structural+magnetic) symmetries of the bulk, the (001) surface of such non-centrosymmetic structures also acquire a unique set of symmetries. This is something not discussed in much detail in the literature and are, in particular, crucial to understand the topologically protected surface physics of these class of compounds. We have described the same in the Appendix \ref{app1}. In the presence of bulk altermagnetic order, it was shown in section \ref{bulk} that $\mathcal{T}$ symmetry is broken. Thus, bulk altermagnetic order may also lead to spin-splitting of the surface bands although they may not \textit{alternate} as explained in the appendix. Indeed, the two concentric surface electron pockets around the $\overline{\Gamma}$-point in the 2D FS are a result of spin-splitting due to bulk altermagnetic order and come from separate spin-up and spin-down surface bands. Also, all the surface bands visible along the $\overline{\Gamma}\rightarrow\overline{\text{X}}$ direction in the surface dispersion are spin-split. This highlights the interplay of bulk altermagnetism and non-trivial topology at the surface. More details on this interplay is presented in Fig. S6 and associated discussions in section S2 of SM \cite{supplement}.
To the best of our knowledge, the AFM phase of any compound in the RAl$X$ ($X$=Si,Ge) family has never been simulated in the existing literature, though few of them are experimentally confirmed to show the same. This is the first study to simulate the bulk as well as surface properties in the collinear AFM phase (magnetic ground state of GdAlSi).

\subsection{VUV ARPES} \label{ARPES}

Figure \ref{fig:fig6}(a) displays the unsymmetrized Fermi surface derived from low-energy (VUV) ARPES measurements at  24 K, employing a photon energy of $h\nu$ = 50 eV. Figures \ref{fig:fig6}(c) and \ref{fig:fig6}(e) depict the constant binding energy contours obtained at $E-E_F=-200$ meV and $-500$ meV respectively for the unsymmetrized case. Remarkably, the observed features in the FS, such as (i) the presence of an electron pocket at $\overline{\Gamma}$, (ii) the existence of Fermi arc-like features along the $\overline{\Gamma}-\overline{\text{M}}$ direction (diagonal) which become more pronounced at deeper energies (Fig. \ref{fig:fig6}(c,e)), (iii) and a fish tail-like pattern at $\overline{\text{X}}$ and a similar state at $\overline{\text{Y}}$, closely resemble the features observed in the calculated Fermi surface (see Fig. \ref{fig:fig6}(b)). To better understand these features, energy-momentum cuts are taken along the $\overline{\Gamma}-\overline{\text{X}}$ direction (see Figs. S7 and S8 and associated discussions in section S3 of SM \cite{supplement}). 

Figure \ref{fig:fig3}(a) displays the intricate features observed in the band dispersion at a photon energy of 40 eV (measured at 24 K). To make the features sharper and more prominent, a second derivative image analysis was performed on Fig. \ref{fig:fig3}(a), the result of which is shown in Fig. \ref{fig:fig3}(b). The energy and momentum scale in the ARPES results (Fig. \ref{fig:fig3}(a,b)) and the calculated results (Fig. \ref{fig:fig3}(c)) are kept the same to facilitate direct comparison between the measured and the simulated dispersions.
To distinguish between the contributions of surface and bulk bands, photon energy dependent ARPES measurements are carried out over the photon energy range 40 to 60 eV at a temperature of 24 K, as depicted in Fig. \ref{fig:fig3}(d). Notably, the ARPES measurements reveal the presence of various features, such as (i) an electron pocket-like state at $\overline{\Gamma}$ (ii) a crossing-like band feature at $\sim$ -1.4 eV near the edge of the bulk-energy gap at $\overline{\Gamma}$ (iii) two downward parabolic band dispersions at deeper energies ($\sim$ -2.0 eV and -2.5 eV, respectively) at $\overline{\Gamma}$ (iv) an upward parabolic-like band dispersion around $\sim$ -0.9 eV at $\overline{\text{X}}$, and (v) a pronounced band dispersion around $\sim$ -2.4 eV at $\overline{\text{X}}$. These features remain largely unchanged in terms of energy and position with respect to photon energy (see Fig. \ref{fig:fig3}), confirming their origin as dominant surface states. However, some other dispersion features in close proximity to the $E_F$ around the $\overline{\text{X}}$ point turn out to be sensitive to the photon energy, indicating their bulk nature. These observed features manifest as a continuum, providing further evidence of bulk states, which is also supported by our calculations.
It is to be noted that, as our ARPES experiment in the VUV range is quite surface sensitive, the results are dominated by both trivial and non-trivial surface states. To probe the bulk \textit{altermagnetic} band dispersion, a soft X-ray ARPES experiment is needed, which we plan to conduct it in the near future.

\subsection{Device models} \label{device}

\begin{figure*}[!htb]
    \centering
    \includegraphics[width=0.8\linewidth]{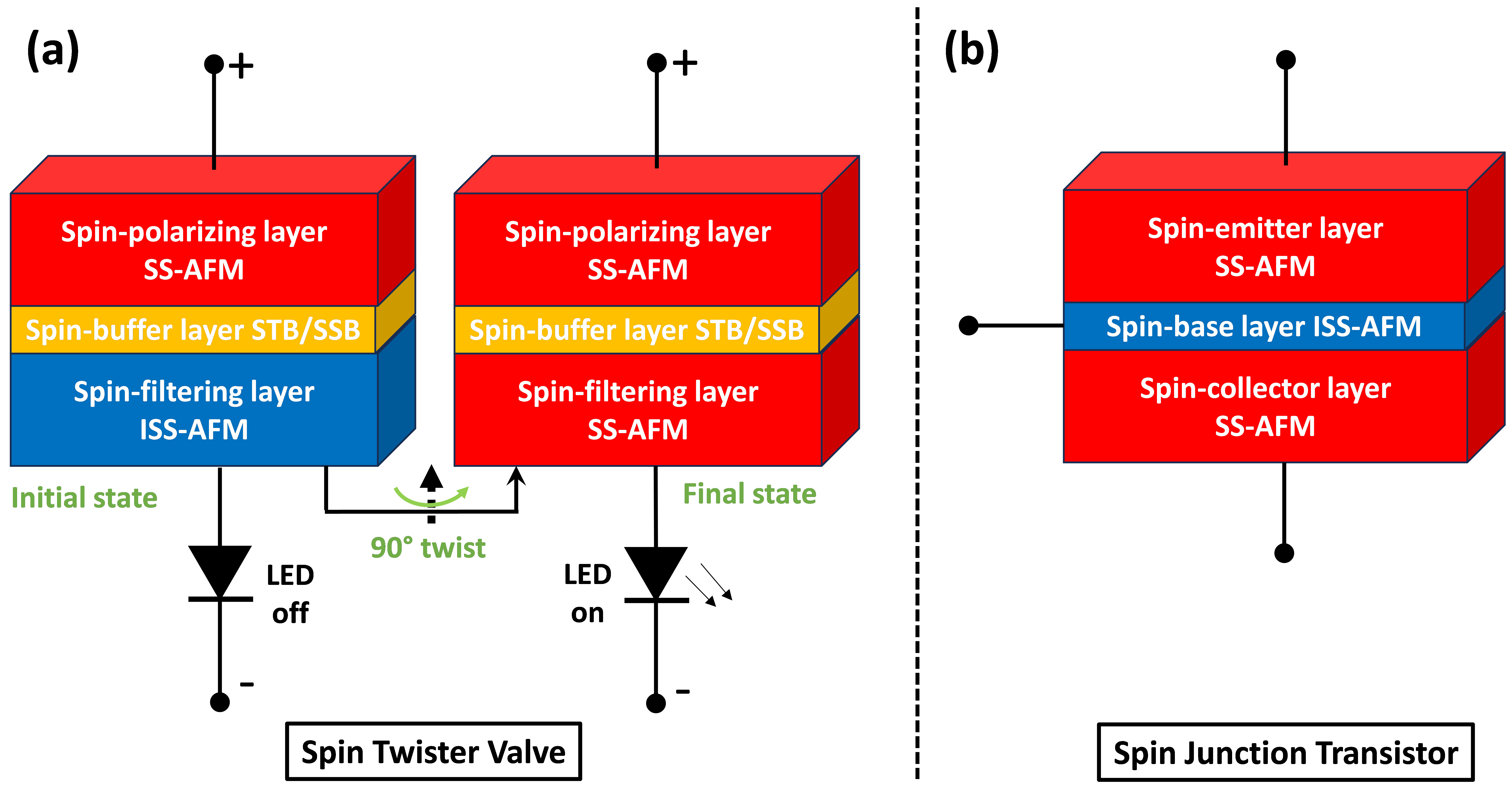}
    \caption {Proposed device models: (a) a spin twister valve and (b) a spin junction transistor utilizing GdAlSi as the active layer. SS-AFM, ISS-AFM and STB/SSB stands for Spin-split Antiferromagnet (GdAlSi in this case), Inverted Spin-split Antiferromagnet (inverted GdSiAl in this case) andSwitching Tunnel Barrier/Spin-sensitive Barrier respectively. LED stands for Light-emitting Diode.}
    \label{fig:fig5}
\end{figure*}

It is clearly evident that the spin-polarization in GdAlSi is highly path-dependent and can be flipped by swapping the Al and Si environments around the inequivalent Gd atoms (see Sec. \ref{bulk} and \ref{nrss}). This unusual feature can have interesting consequences useful for practical devices.

In case of RuO$_2$ (another candidate for altermagnetism, but centrosymmetric), it is shown that application of electric field along specific crystal directions can result in generation of highly spin-polarized current \cite{Gonzalez2021,Bai2022}. A similar response is expected in GdAlSi due to the spin anisotropy in the bulk BZ. Let us consider the Al and Si environments around the Gd atoms to have been swapped as illustrated in section \ref{nrss} such that GdAlSi now becomes the hypothetical GdSiAl. Since swapping the non-magnetic environment results in flipping of spin-polarized bands (see Figs. \ref{fig1}(a),(c),(e),(g)), it is expected that the spin-polarization in GdAlSi and GdSiAl will be opposite for a given electric field applied along a certain crystal direction.

Another interesting feature of GdAlSi and GdSiAl structures is their connection via simple crystal transformations. If one inverts the GdSiAl structure along the $c$ axis (i.e., $c\rightarrow-c$) and gives a $C_{4z}$ transformation (rotation by 90$^{\circ}$ about the inverted $c$ axis) to it, one gets back the original GdAlSi structure. Our ab initio calculations show that the relative energy difference between GdAlSi and GdSiAl structures is $\sim$ 1 meV/atom. This implies that growing the hypothetical ordered GdSiAl is practically feasible.

Now if a GdAlSi layer and an \textit{inverted} GdSiAl layer are sandwiched together with a spin current sensitive layer in between, one obtains what we call a spin twister valve (STV). A device architecture for the same is shown in Fig. \ref{fig:fig5}(a). Here, \textit{inverted} refers to inverted growth of GdSiAl layer which we believe can be achieved with current state-of-art crystal/film growth techniques. The top layer would act as a \textit{spin polarizing layer} and the bottom layer as a \textit{spin filtering layer}. The middle layer acts as a \textit{spin buffer layer} which transfers the spin-polarized current from the top to the bottom layer. This may be done in two ways. One is via a non-magnetic dielectric layer acting as a switching tunnel barrier (STB) similar to those found in magnetic tunnel junctions (MTJ). The other is via a topological insulator layer with spin-polarized surface states acting as a spin-sensitive barrier (SSB). Since GdAlSi is itself a topological WSM with spin-polarized Fermi arcs as explained in section \ref{surface}, this would enable surface coupling of these spin-polarized states leading to ballistic spin channels and ultra-high mobility. The functioning of the device relies on the opposite spin polarizations of the top GdAlSi and the bottom \textit{inverted} GdSiAl layer. When an electric field is applied across the device, initially if the top layer generates a spin-up current, then the bottom layer completely blocks it (LED off state as shown in left circuit of Fig. \ref{fig:fig5}(a)). Now if the bottom layer is \textit{mechanically twisted} by 90$^{\circ}$, it is transformed into the top layer and the spin-up current can pass through the device (LED on state as shown in right of Fig. \ref{fig:fig5}(a)).

Another device architecture is possible by removing the middle layer in the proposed STV and adding another GdAlSi layer at the bottom. This would form what we call a spin junction transistor (SJT) as shown in Fig. \ref{fig:fig5}(b). This would be a three-terminal device and function analogous to the semiconductor-based bipolar junction transistors (BJT) but with spin currents. The top and the bottom GdAlSi layers would behave as the \textit{spin emitter layer} and the \textit{spin collector layer} respectively. The sandwiched \textit{inverted} GdSiAl layer would behave as the \textit{spin base layer}. In a SJT, the majority and minority carriers would be spin-up and spin-down electrons (or vice-versa) respectively unlike in BJTs where the majority and minority carriers are electrons and holes (or vice-versa) respectively. The junction between the layers would act as \textit{spin depletion layers} akin to depletion layers in semiconductor junctions due to the possibility of spin diffusion among the layers. When appropriately biased, one would expect SJTs to behave similar to BJTs but strictly with spin-polarized currents. The operation of such SJTs could have greater efficiency due to the presence of topological spin polarized Fermi arcs of WSMs providing spin ballistic channels. Similar to transistors which are well-known to bring technological revolution since their invention, SJTs could provide the much-awaited breakthroughs in Spintronics.

We would like to emphasize that these device architectures are entirely generic and can be extended to any spin-split antiferromagnets (SS-AFM), which in the present case is GdAlSi.

\section{Conclusions}

In summary, we present a comprehensive study of the electronic, magnetic and topological properties of GdAlSi employing a combined {\it{ab-initio}} calculations, SHG, magneto-transport and VUV ARPES measurements. This is the first non-centrosymmetric ($I4_{1}md$ ($109$)) system which shows the coexistence of wondrous \textit{altermagnetism} feature and antiferromagnetic Weyl semimetal behavior. This is confirmed via a rigorous symmetry analysis, electric/magnetic multipole analysis, ARPES measurements and detailed structural analysis through XRD and optical SHG experiments. Magnetic and transport measurements suggest GdAlSi to exhibit AFM behavior below 32 K. First-principles calculations and VUV ARPES measurements confirm the presence of Fermi arc-like features, providing strong evidence of the Weyl semimetallic state. The theoretical predictions are in fair agreement with the experimental findings, further supporting the existence of Weyl points and Fermi arc-like features.

Our theoretical multipole analysis suggest the presence of {\it{altermagnetism}}, as indicated by significant band splitting along a specific path in the Brillouin zone. Altermagnetism has never been predicted before in rare earth materials, especially in the RAlSi family and we believe this to be the first study facilitating the same. 
GdAlSi has the potential to be the next promising topological/spintronic material as it offers a remarkable platform to investigate the Weyl Physics in the presence of \textit{altermagnetism} and non-trivial band topology. The presence of magnetic octupoles in GdAlSi also allows for interesting physical effects such as piezomagnetism, and unconventional magnetic Compton profile in a compensated antiferromagnet \cite{Bhowal2024}. The latter also provides a means for directly detecting magnetic octupoles and the corresponding non-relativistic spin-splitting, motivating future investigations along these directions. Further, we propose device models to effectively utilize this unique coexistence which can provide a new route to practical topotronic devices. Consequently, our study opens up new avenues for exploring the non-trivial band topology combined with altermagnetism with unique practical applications.

\begin{acknowledgements}
    JN acknowledges the financial support, in the form of fellowship, from IIT Bombay and MEXT Japan for MEXT fellowship 2021 (210035) for pursuing research in Hiroshima University, Japan. JN and VG acknowledge the Air Force office of scientific research under award number  FA 9550-23-1-0658 for the optical SHG work in this study. BD acknowledges MHRD India and IIT Bombay for financial support. SB thanks ETH Zurich for financial support. BD thanks Dr. Chanchal K. Barman for stimulating discussions and valuable suggestions.
\end{acknowledgements}

\let\origsection=\section
\let\section=\subsection
\section*{Appendix A: Symmetry Analysis of (001) Surface} \label{app1}

The cleaving of bulk GdAlSi and exposing the (001) surface leads to the reduction of the parent tetragonal space group $I4_{1}md\;(109)$ of bulk GdAlSi to an orthorhombic space group $Pmm2\;(25)$ at the (001) surface. It is seen that the reduced point group $mm2\;(C_{2v})$ on the (001) surface is actually the halving subgroup of the bulk point group $4mm\;(C_{4v})$ as discussed in section \ref{bulk}. But after symmetry reduction due to cleaving of the bulk system, $C_{2v}$ becomes the parent point group and it does not contain the $C_{4z}$ symmetry element. Considering the full space group symmetries, the non-symmorphic translation $\textbf{t}$ will also be absent at the (001) surface (which lies on the crystalline $ab$-plane) since it has a non-zero component along the crystalline $c$-axis. Thus the symmetry transformation $C_{4z}\textbf{t}$ crucial for bulk altermagnetism is absent on the (001) surface of GdAlSi. Although halving subgroups exist for $C_{2v}$, there exists no opposite-spin sublattice transformation at the (001) surface since the opposite-spin sublattices are stacked along the crystalline $c$-axis when bulk AFM-I ordering is considered (see Figs. \ref{fig:fig1}(a) and \ref{fig:fig1}(b)). Hence, dispersion at the cleaved (001) surface of GdAlSi with bulk-like AFM-I order in the absence of SOC is not expected to \textit{alternate} as in the bulk case. 

In the presence of SOC and AFM-I order with the magnetic moments on Gd oriented along the crystalline $c$-axis, the magnetic space group $I4_{1}'m'd$ of GdAlSi is reduced to $Pm'm'2$ at the (001) surface. After symmetry reduction due to cleaving of the bulk system, since $m'm'2$ becomes the parent point group and contains the $C_{2z}$ symmetry element, the 2D FS shows a two-fold symmetry as mentioned earlier. The point group $m'm'2$ also contains the anti-unitary mirror symmetries $\Tilde{\mathcal{M}}_{x} = \mathcal{M}_{x}\mathcal{T}$ and $\Tilde{\mathcal{M}}_{y} = \mathcal{M}_{y}\mathcal{T}$ and hence the 2D FS is symmetric about the $\overline{\Gamma}\rightarrow\overline{\text{X}}$ and $\overline{\Gamma}\rightarrow\overline{\text{Y}}$ directions individually. The distinct asymmetry along $\overline{\text{X}}$ and $\overline{\text{Y}}$ as mentioned earlier is due to the absence of the mirror symmetries $\mathcal{M}_{xy}$ and $\mathcal{M}_{x\overline{y}}$ in $m'm'2$ which were present in the bulk point group $4'm'm$. We note here that compounds that crystallize in the $I4_{1}md$ space group without any magnetic order will have the magnetic point group $mm21'$ on the (001) surface in the presence of SOC. Since $mm21'$ also does not contain the $\mathcal{M}_{xy}$ and $\mathcal{M}_{x\overline{y}}$ symmetries, the corresponding 2D FS will again be asymmetric as seen in TaAs \cite{xu2015discovery} and LaAlGe \cite{xu2017discovery}. Previous literature on ferromagnetic RAl$X$ ($X$=Si,Ge) family of compounds such as PrAlGe have attributed this observed FS asymmetry to broken $\mathcal{P}$ symmetry and the breaking of $\mathcal{T}$ symmetry by the underlying ferromagnetic order in PrAlGe \cite{sanchez2020observation}.  Interestingly, we find that for magnetic moments oriented along the crystalline $c$-axis, the point group at the (001) surface is again $m'm'2$ which allows for ferromagnetic order. Hence, the aforementioned asymmetry is actually due to the absence of $C_{4z}$, $\mathcal{M}_{xy}$ and $\mathcal{M}_{x\overline{y}}$ symmetries as we explained earlier. Such ferromagnetic order can lead to spin-polarization and spin-splitting of surface bands which has not been highlighted much in previous literature \cite{sanchez2020observation, PhysRevMaterials.7.L051202} but may prove to be useful in spintronic device applications as explained later in section \ref{device}. \\
\let\section=\origsection

\bibliographystyle{revtex4-2.bst}
\bibliography{main}
\clearpage


\title{Supplementary Material for `` GdAlSi: An antiferromagnetic topological Weyl semimetal with non-relativistic spin splitting "}

\maketitle

\onecolumngrid
\setcounter{section}{0}
\setcounter{equation}{0}
\setcounter{figure}{0}
\setcounter{table}{0}
\makeatletter
\renewcommand{\bibnumfmt}[1]{[S#1]}
\renewcommand{\citenumfont}[1]{S#1}
\renewcommand{\thesection}{S\arabic{section}}
\renewcommand{\thefigure}{S\arabic{figure}}
\renewcommand{\thetable}{S\arabic{table}}

\begin{bibunit}

    \noindent Here, we present further auxiliary details of experimental synthesis, different measurement tools, and \textit{ab-initio} computation. We have demonstrated the details of the Laue, XRD refinement, EDS, SHG data analysis, and resistivity for GdAlSi. We also present further results on \textit{ab-initio} calculations along with the ARPES data.
    
    \section{Structure characterization}
    
    \subsection{Elemental Analysis}
\begin{figure*}[!htb]
        \centering
        \includegraphics[width=1.0\linewidth]{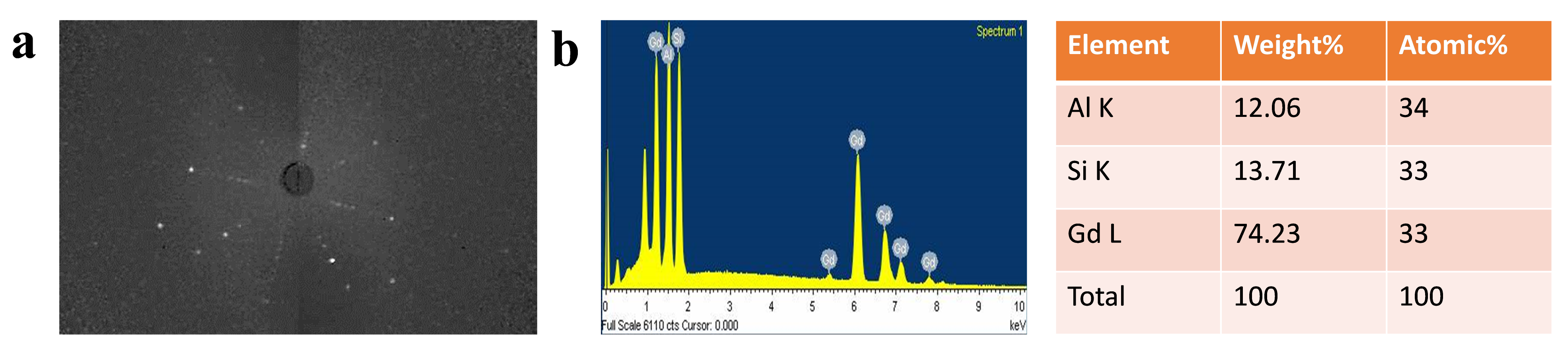}
        \caption{For GdAlSi (a) Laue pattern (b) composition analysis on a single crystal and weight/atomic percent of different constituent elements (right table).}
        \label{fig:fig01_8}
    \end{figure*}

    Multiple single crystals of GdAlSi were synthesized and carefully examined for elemental analysis. The rare-earth standards used in the analysis exhibited partial surface degradation, leading to a relatively broad range of total composition, approximately ranging from 97\%-102\%. However, despite this variation, the relative quantities of Al and Si were accurately determined. The resulting Al:Si ratios were found to be nearly 50:50 in all the samples. After conducting multiple compositional analysis, nearly 1:1:1 ratio of Gd:Al:Si  were obtained by averaging the results, with normalization per rare-earth metal (see Fig. \ref{fig:fig01_8}).
    
    \vspace*{-1cm}
    \subsection{Structural Properties}
    Laue pattern, as shown in  Fig. \ref{fig:fig01_8}(a), confirms the high quality of the single crystal with tetragonal symmetry. The Rietveld refined atomic coordinates are presented in Table \ref{table:wyckoff}.
     
    \subsection{SHG Polarimetry} 
    To validate the noncentrosymmetric space group $I4_{1}md$ (point group \( 4mm \)) classification established through single crystal X-ray refinement, optical second harmonic generation experiment was performed using reflection geometry at 45$^o$ oblique incidence with a 800 nm fundamental laser beam (Spectra-Physics Ti:sapphire source, 80 fs, 80 MHz). Figure 1(c) of the main manuscript shows the schematic of the SHG setup used. Subsequently, the reflected SHG field was separated into $p$-polarized (X) and $s$-polarized (Y) components using an analyzer and detected by a photomultiplier tube. For SHG polarimetry, nonlinear optical intensity as a function of the incidence angle of the linearly polarized light has been measured.
    The resulting polar plots were fitted based on point group $4mm$, utilizing the \#SHAARP open-source package \cite{paper2} to simulate the second harmonic response of anisotropic materials. 
     
    \subsection{SHG Results}
    The SHG process in a nonlinear optical material creates an induced nonlinear polarization, $P^{2 \omega}$ at frequency $2 \omega$ in response to an incoming electric field $E^{\omega}$ of light of frequency $\omega$, which are related by the second-order optical susceptibility $P_i^{2 \omega}=d_{i j k} E_j^{\omega} E_k^{\omega}$ \cite{paperr}. Here, the indices $(i,j,k)$ are dummy variables denoting the polarization directions in Cartesian coordinates. SHG measurements were performed on single crystals of GdAlSi.
    Figure 1(d) of the manuscript presents the measured SHG polar plots as a function of the incident polarization angle of the fundamental beam from a GdAlSi single crystal, indicating the non-centrosymmetric nature of the crystal. The second harmonic tensor for point group $4mm$ is represented in Voigt notation as: 
    \begin{equation}
        d=\begin{pmatrix} 
        0 & 0 & 0 & 0 & d_{15}& 0 \\
        0 & 0 & 0 &  d_{15}& 0 & 0 \\
        d_{31} & d_{31} & d_{33} & 0 &  0  & 0 \\
    \end{pmatrix}   
    \end{equation}
    Here $d_{15}$, $d_{31}$ and $d_{33}$  are three non-zero SHG coefficients. The SHG intensity, $I^{2\omega}$, at 45$^o$ oblique incidence can be derived using \#SHAARP  \cite{paper2} as:
    \begin{equation}
    \begin{split}
      I_s^{2\omega} \propto   \lvert ((2d_{15}-d_{31}-d_{33})\cos^{2}{\phi}-2d_{31}\sin^{2}{\phi})^2\rvert \\
      I_p^{2\omega} \propto   \lvert (2d_{15}^2\cos^{2}{\phi}\sin^{2}{\phi})\rvert
      \end{split}
      \label{eq:02}
    \end{equation}
    The experiment was conducted using a rotating polarizer and fixed analyzer configuration, and the subscripts $s$ and $p$ respectively represent the \textit{s}-polarization and the \textit{p}-polarization as shown in Fig. 1(c) of the main manuscript. Experimentally, the intensity of second harmonic as a function fundamental input intensity has been measured (see Fig. 1(e) of the main manuscript), which shows a quadratic dependence, namely, $I^{2\omega}$= A*{$I^\omega$}$^2$ establishing the lack of inversion symmetry of the GdAlSi single crystals. The combined result from XRD refinement and SHG polarimetry measurement at 45$^o$ oblique incidence (reflection geometry) demonstrates an excellent agreement with the theory in Eq. \ref{eq:02}, thus establishing the crystals to have a noncentrosymmetric $4mm$ point group. 
    The combined results from XRD refinement and SHG experiments confirm that GdAlSi crystallizes in the non-centrosymmetric space group $I4_{1}md$ (point group \( 4mm \)).

    \begin{table*}
        \centering
        \caption {\textcolor{black}{Experimental and theoretically optimized lattice parameters along with} Rietveld refined atomic coordinates of bulk GdAlSi }
        \begin{center}
        \begin{tabular}{|c|c|c|c|c|c|c|c|c|}
        \hline
        \multicolumn{4}{|c|}{\textbf{Lattice parameters} (\AA)} & \multirow{3}{*}{\textbf{Atom}} & \multirow{3}{*}{\shortstack{\textbf{Wyckoff} \\ \textbf{position}}} & \multicolumn{3}{c|}{\multirow{2}{*}{\shortstack{\textbf{Atomic} \\ \textbf{coordinates}}}} \\
        \cline{1-4}
        \multicolumn{2}{|c|}{\textbf{Experiment}} & \multicolumn{2}{c|}{\textbf{Simulated}} & & & \multicolumn{3}{c|}{} \\
        \cline{1-4} \cline{7-9}
        $a=b$ & $c$ & $a=b$ & $c$ & & & \textbf{$x$} & \textbf{$y$} & \textbf{$z$} \\
        \hline
        \cline{1-7}
         \multirow{5}{*}{4.12} & \multirow{5}{*}{14.43} & \multirow{5}{*}{4.14} & \multirow{5}{*}{14.50} & & & & & \\
          & & & & Al  &  4a  &  0.0 &  0.0 &  0.958(2) \\
          & & & & Gd  &  4a  &  0.0 &  0.0 &  0.374(2) \\
          & & & & Si  &  4a  &  0.0 &  0.0 &  0.793(2) \\
          & & & & & & & & \\
        \hline
        \end{tabular}
        \end{center}
        \label{table:wyckoff}
    \end{table*}
    
    \subsection{Resistivity}
    Figure 1(g) of the manuscript displays the temperature-dependence of resistivity at two different magnetic fields. A distinct kink at/around 32 K confirms the antiferromagnetic (AFM) transition, which is consistent with previous magnetic and resistivity measurements \cite{bobev2005ternary_sm}.

    \section{\textit{Ab-initio} results} \label{theory}
    
    \subsection{Ground state magnetic structure}
\begin{figure*}[!p]
        \centering
        \includegraphics[width=0.85\linewidth]{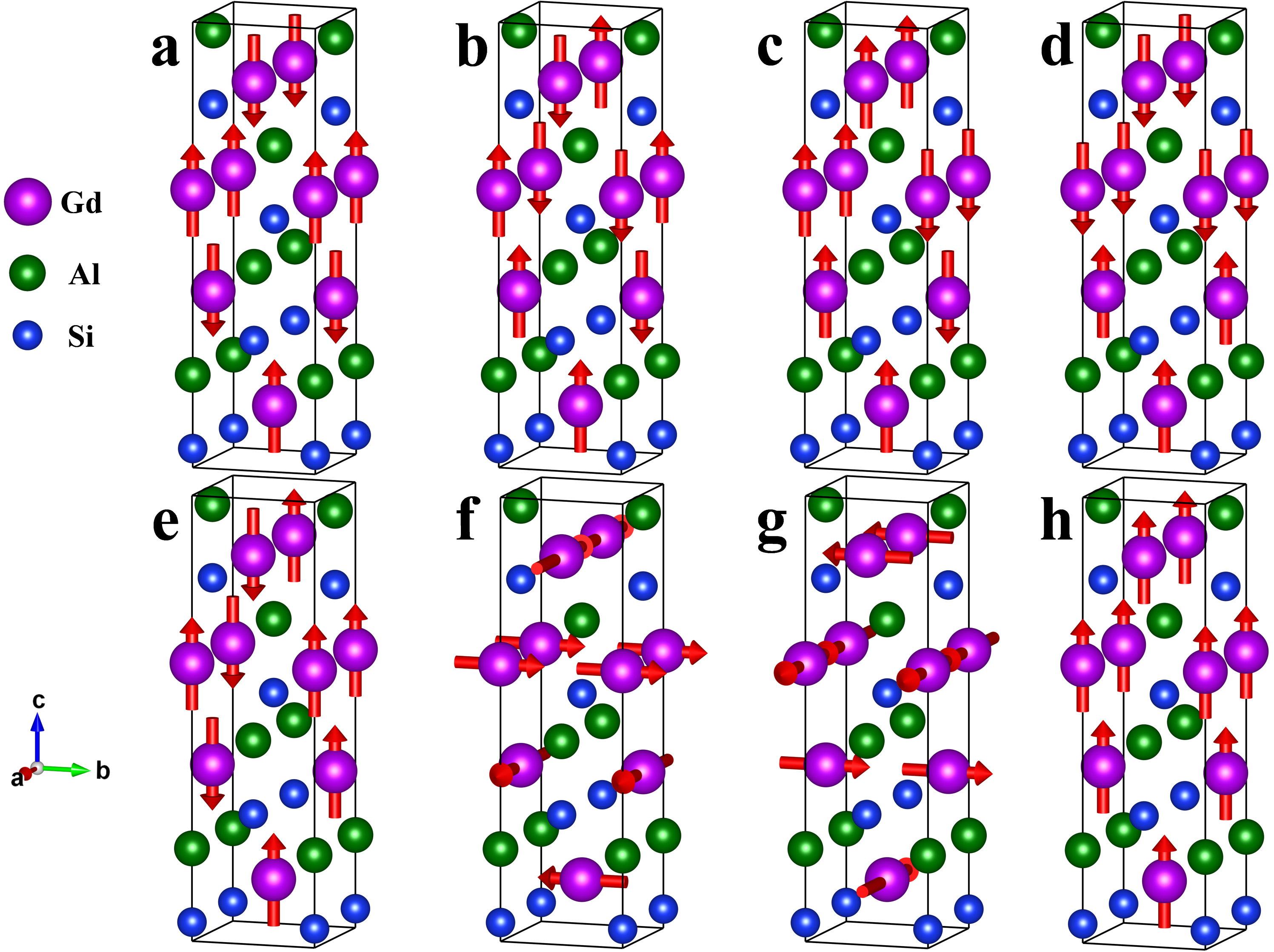}
        \caption{Different magnetic configurations of GdAlSi namely (a) AFM-I, (b) AFM-II, (c) AFM-III, (d) AFM-IV, (e) FiM, (f) Helical ($+$), (g) Helical ($-$) and (h) FM.}
        \label{mag_struct}
    \end{figure*}

    First principle calculations were performed on GdAlSi considering different magnetic configurations as shown in Fig. \ref{mag_struct}. \textcolor{black}{Recently, NdAlSi has been shown to possess a commensurate ferrimagnetic (FiM) order \cite{gaudet2021weyl_sm,li2023_sm} which has also been simulated here for GdAlSi. Among non-collinear configurations, a period-4 helical magnetic order with the magnetic moments oriented on the crystalline $ab$-plane has also been simulated. Both right-handed and left-handed helical magnetic order have been considered, denoted by ``Helical ($+$)" and ``Helical ($-$)" respectively.} The simulated relative energies and Gd-projected moments for each configuration are summarized in Table \ref{table}. Out of all the configurations, AFM-I turns out to be the lowest in energy.
    
    \begin{table*}[!p]
        \caption {Simulated relative total energies and Gd-projected local moments for different magnetic configurations of GdAlSi}
        \label{table}
        \begin{center}
        \begin{tabular}{|c|c|c|}
        \hline
        \multirow{3}{*}{\shortstack{\textbf{Magnetic}\\\textbf{configuration}}} & \multirow{3}{*}{\shortstack{\textbf{Relative total energy}\\\textbf{(meV)}}} & \multirow{3}{*}{\shortstack{\textbf{Local magnetic moment}\\\textbf{on Gd atom ($\mu_B$)}}} \\
        & & \\
        & & \\
        \hline
        \multirow{2}{*}{AFM-I} & \multirow{2}{*}{0} & \multirow{2}{*}{7.087}\\
        & & \\
        \multirow{2}{*}{AFM-II} & \multirow{2}{*}{29} & \multirow{2}{*}{7.083} \\
        & & \\
        \multirow{2}{*}{AFM-III} & \multirow{2}{*}{53} & \multirow{2}{*}{7.076} \\
        & & \\
        \multirow{2}{*}{AFM-IV} & \multirow{2}{*}{137} & \multirow{2}{*}{7.057} \\
        & & \\
        \multirow{2}{*}{FiM} & \multirow{2}{*}{82} & \multirow{2}{*}{7.068} \\
        & & \\
        \multirow{2}{*}{Helical ($+$)} & \multirow{2}{*}{147} & \multirow{2}{*}{7.055} \\
        & & \\
        \multirow{2}{*}{Helical ($-$)} & \multirow{2}{*}{147} & \multirow{2}{*}{7.055} \\
        & & \\
        \multirow{2}{*}{FM} & \multirow{2}{*}{327} & \multirow{2}{*}{7.012} \\
        & & \\
        \hline
        \end{tabular}
        \end{center}
    \end{table*}

    \subsection{Spin-polarized bulk Fermi surface}
\begin{figure}[!ht]
        \centering
        \includegraphics[width=1.0\linewidth]{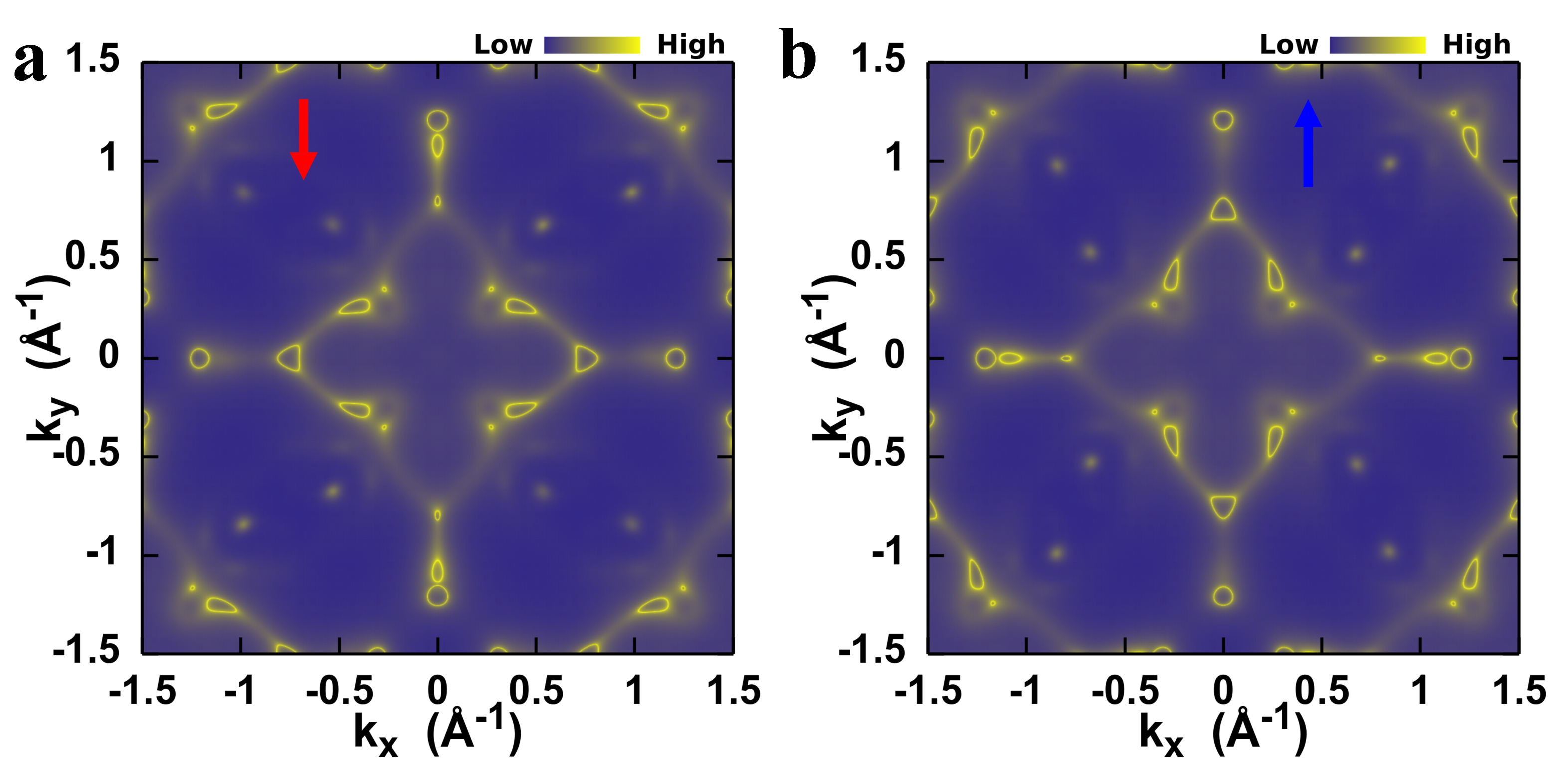}
        \caption{Projection of the bulk Fermi surface of GdAlSi in AFM-I magnetic configuration on $k_z=0$ plane for (a) down spin (red down arrow) and (b) up spin (blue up arrow).}
        \label{spin_FS}
    \end{figure}

    In the absence of spin-orbit coupling, the bulk Fermi surface (FS) of GdAlSi in the AFM-I configuration is spin-split which has been attributed to the presence of altermagnetism as discussed in detail in Sec. III(C) of the main manuscript. The projection of the bulk FS on the $k_z = 0$ plane is shown in Fig. \ref{spin_FS}. Interestingly, the bulk FS is not only spin-split but the spin up and spin down FS are rotated by 90$^{\circ}$ with respect to each other as a consequence of the opposite spin sublattice transformation $[C_2||C_{4z}\textbf{t}]$.

    \begin{figure}[p]
        \centering
        \vspace{-0.55cm}
        \includegraphics[width=1.0\linewidth]{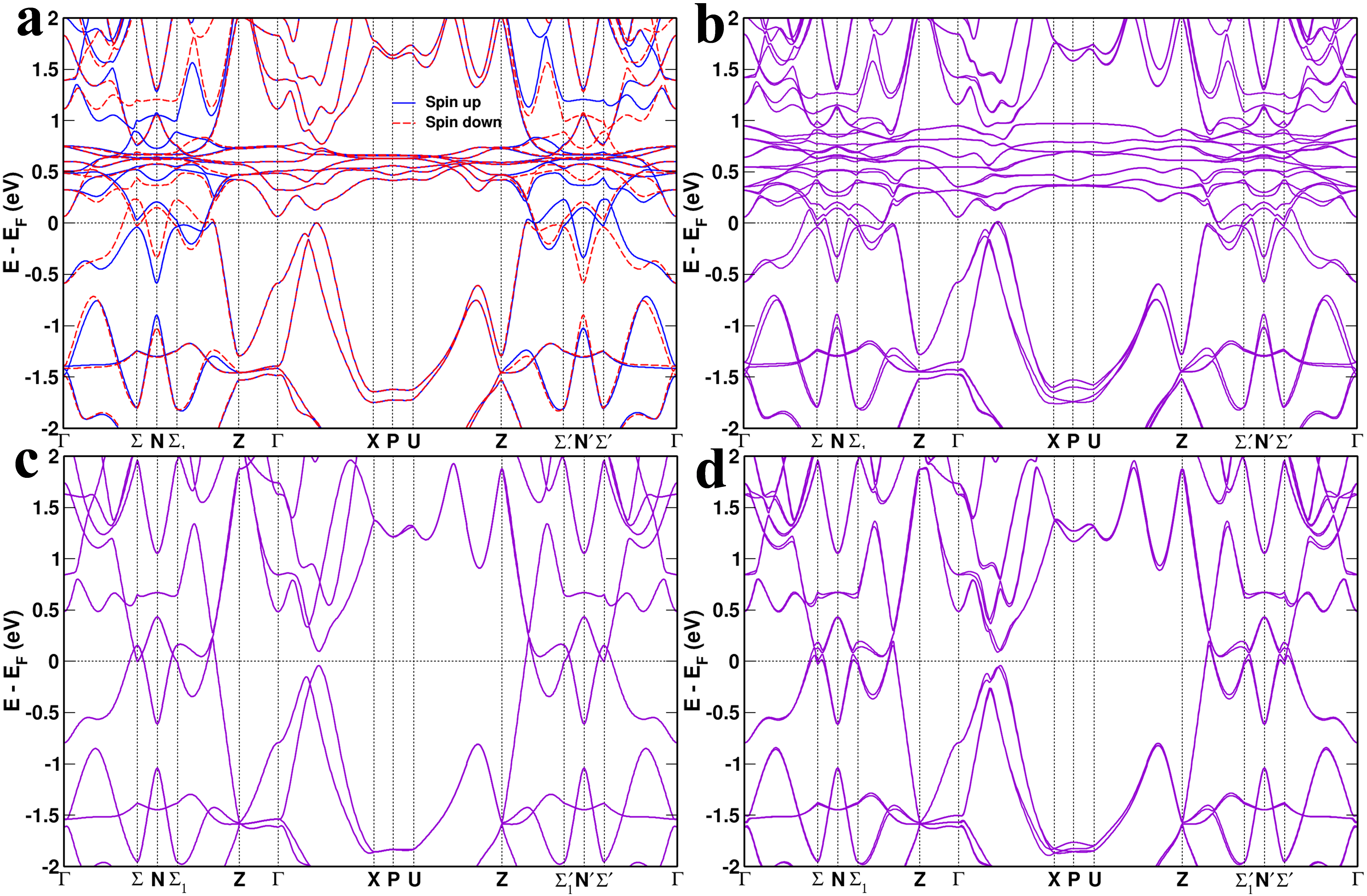}
        \caption{For GdAlSi, bulk band structure (a) without, (b) with SOC for AFM-I configuration (U = 0), (c) without, (d) with SOC for paramagnetic configuration.}
        \label{fig:figs4}
    \end{figure}

    \begin{figure}[p]
        \centering
        \vspace{-0.2cm}
        \includegraphics[height=9.8cm,width=17cm]{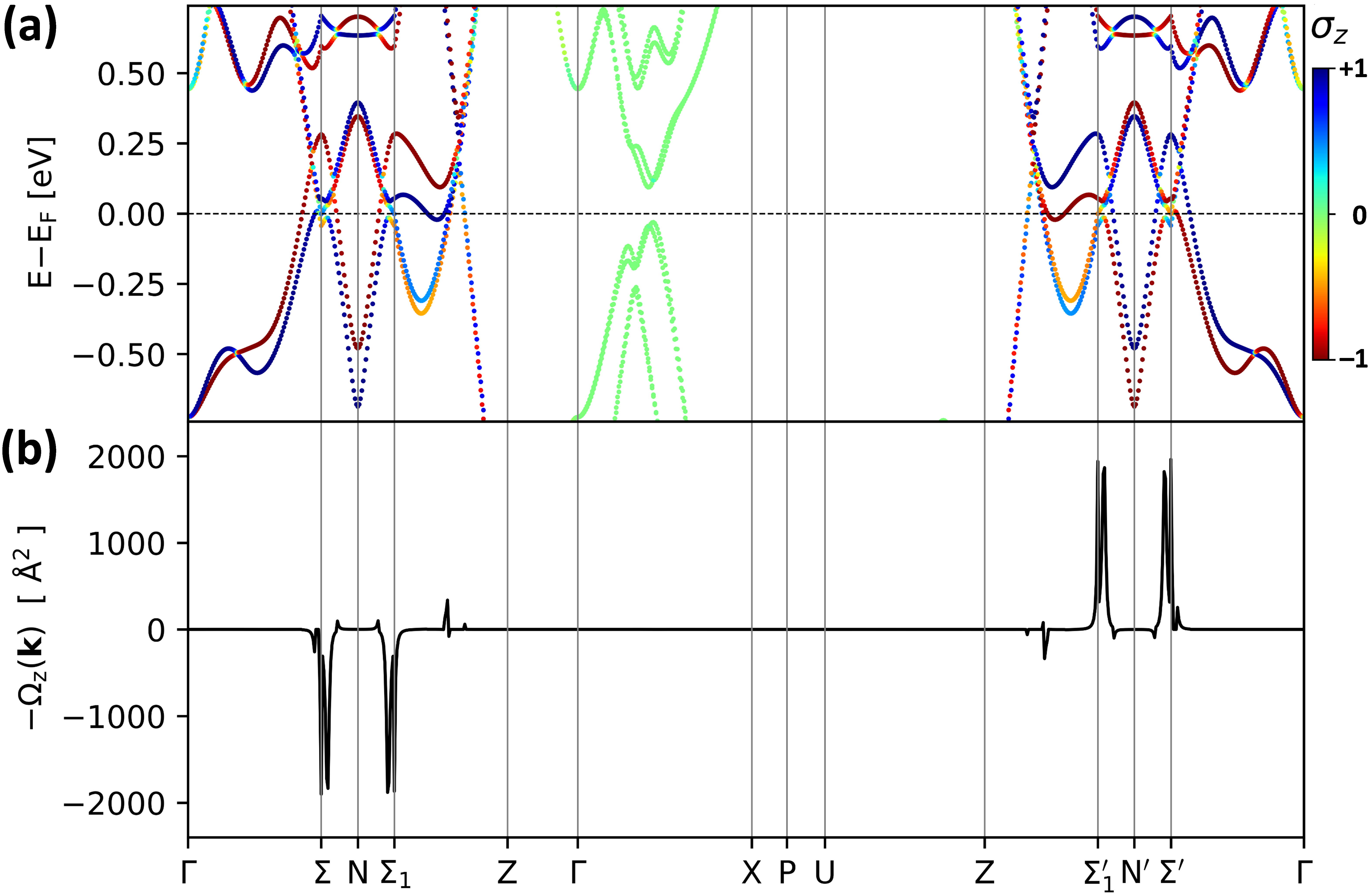}
        \caption{For AFM-I configuration of GdAlSi, (a) simulated SOC bulk band structure with $\sigma_z$ projections (spin components along the $z$-axis) and (b) total Berry curvature calculated at the Fermi level along the same high symmetry $k$-paths.}
        \label{berry}
    \end{figure}
    
    \begin{figure}[p]
        \centering
        \vspace{-1cm}
        \includegraphics[width=0.7\linewidth]{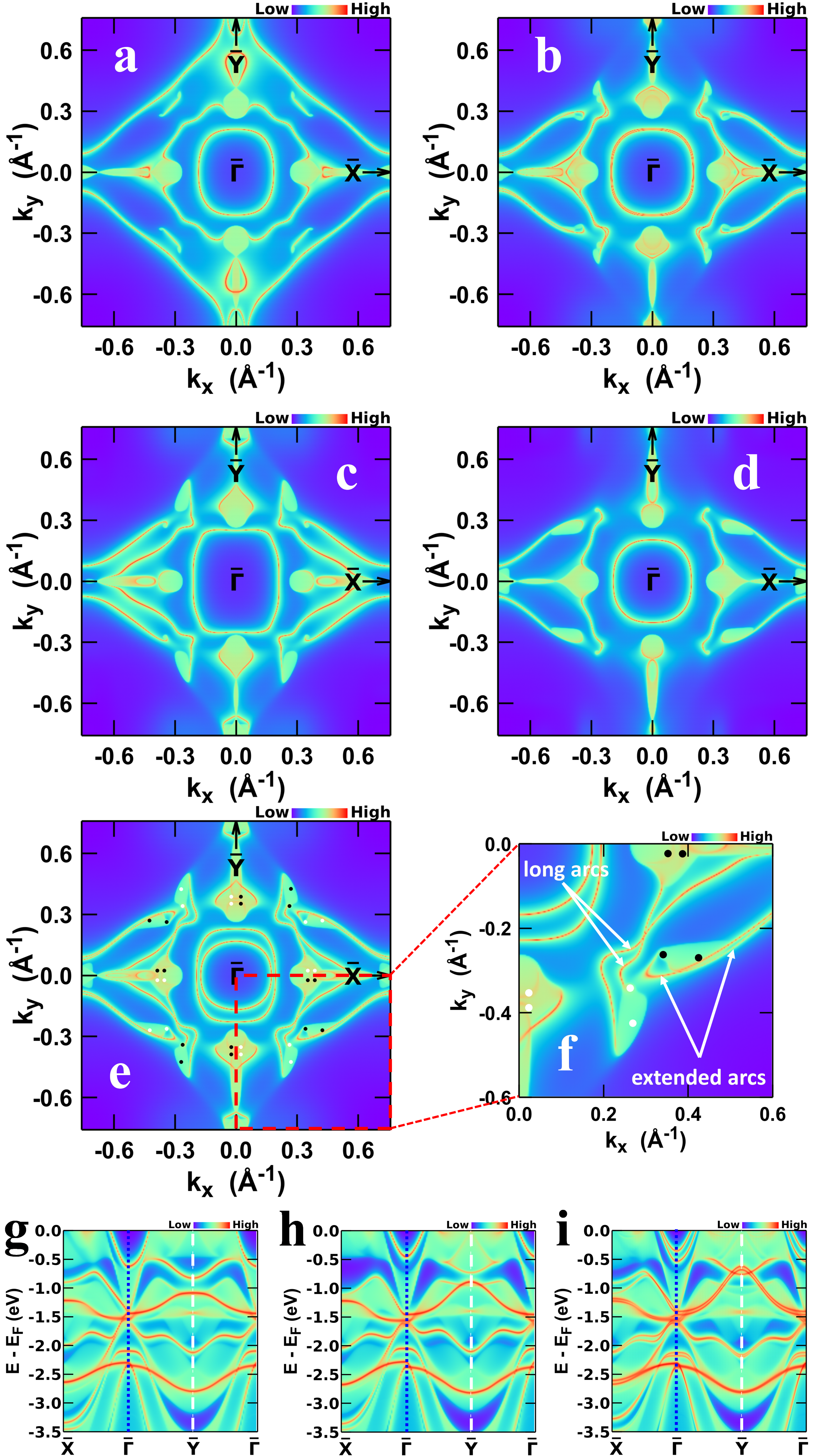}
        \caption{For the (001) surface of GdAlSi, symmetrized simulated Fermi surface for paramagnetic configuration (a) without SOC (b) with SOC. The same for AFM-I configuration without SOC for (c) spin-up and (d) spin-down polarizations. (e) With SOC, the symmetrized simulated Fermi surface for AFM-I configuration at the (001) surface. The white (black) dots denote the surface projections of the bulk Weyl points with +1 (-1) chirality. The red dashed square displays (f) a zoomed-in quadrant of the surface BZ, where the white arrows are the long intra-BZ Fermi arcs and extended inter-BZ Fermi arcs connecting the type-I Weyl points. The surface dispersion for AFM-I configuration without SOC, for (g) spin-up and (h) spin-down polarizations, and (i) with SOC. The vertical blue dotted and white dashed lines locate $\overline{\Gamma}$ and $\overline{\text{Y}}$ points respectively. There is a noticeable asymmetry along the $\overline{\Gamma}-\overline{\text{X}}$ and $\overline{\Gamma}-\overline{\text{Y}}$ directions in the Fermi surfaces as well as the surface dispersions. Further, this asymmetry is different for spin-up and spin-down polarizations due to presence of bulk altermagnetism.}
        \label{fig:figs6}
    \end{figure}
    
    \begin{table*}[p]
        \centering
        \vspace{2.5cm}
        \caption {Reciprocal space coordinates, type, energy position and chirality of 16 pairs of Weyl points (WPs) in the band structure of bulk GdAlSi.}
        \begin{center}
        \begin{tabular}{|c|c|c|c|c|c|c|}
        \hline
        \multirow{3}{*}{\textbf{\#}} & \multicolumn{3}{c|}{\multirow{2}{*}{\shortstack{\textbf{Cartesian coordinates} \\ \textbf{of WPs}}}} & \multirow{3}{*}{\textbf{Type}} & \multirow{3}{*}{\shortstack{\textbf{Relative energy} \\ \textbf{from} $\text{\textbf{E}}_\text{\textbf{F}}$ \textbf{(meV)}}} & \multirow{3}{*}{\textbf{Chirality}} \\
        & \multicolumn{3}{c|}{} & & & \\
        \cline{2-4}
        & \textbf{$k_x$} (\AA$^{-1}$) & \textbf{$k_y$} (\AA$^{-1}$) & \textbf{$k_z$} (\AA$^{-1}$) & & & \\
        \hline
        \cline{1-7}
          & & & & & & \\
          1  &  0.023 &  0.352 & -0.359 & IIA & 58.9 & -1 \\
          2  &  0.024 &  0.387 &  0.288 & IIB & 65.5 & -1 \\
          3  & -0.023 &  0.352 &  0.359 & IIA & 58.2 &  1 \\
          4  & -0.024 &  0.387 & -0.288 & IIB & 65.5 &  1 \\
          5  &  0.024 & -0.387 & -0.288 & IIB & 65.8 &  1 \\
          6  & -0.024 & -0.387 &  0.288 & IIB & 65.6 & -1 \\
          7  & -0.023 & -0.352 & -0.360 & IIA & 58.6 & -1 \\
          8  &  0.024 & -0.352 &  0.360 & IIA & 58.3 &  1 \\
          9  &  0.387 & -0.023 & -0.288 & IIB & 64.6 & -1 \\
         10  &  0.387 &  0.023 &  0.288 & IIB & 64.3 &  1 \\
         11  &  0.352 & -0.023 &  0.359 & IIA & 58.3 & -1 \\
         12  &  0.263 &  0.341 &  0.003 &  IA & 23.9 & -1 \\
         13  &  0.341 &  0.263 &  0.003 &  IA & 24.5 &  1 \\
         14  &  0.270 &  0.425 & -0.003 &  IB & 79.2 & -1 \\
         15  &  0.270 & -0.425 &  0.003 &  IB & 79.2 &  1 \\
         16  &  0.263 & -0.341 & -0.003 &  IA & 23.7 &  1 \\
         17  &  0.341 & -0.263 & -0.003 &  IA & 24.2 & -1 \\
         18  &  0.352 &  0.023 & -0.359 & IIA & 58.4 &  1 \\
         19  & -0.341 & -0.263 &  0.003 &  IA & 24.3 &  1 \\
         20  &  0.425 &  0.269 & -0.003 &  IB & 78.9 &  1 \\
         21  &  0.425 & -0.269 &  0.003 &  IB & 79.0 & -1 \\
         22  & -0.388 &  0.023 & -0.287 & IIB & 64.5 & -1 \\
         23  & -0.425 & -0.269 & -0.003 &  IB & 79.0 &  1 \\
         24  & -0.425 &  0.269 &  0.003 &  IB & 79.0 & -1 \\
         25  & -0.388 & -0.023 &  0.287 & IIB & 64.6 &  1 \\
         26  & -0.341 &  0.263 & -0.003 &  IA & 24.5 & -1 \\
         27  & -0.353 &  0.023 &  0.360 & IIA & 57.3 & -1 \\
         28  & -0.270 & -0.424 & -0.003 &  IB & 79.1 & -1 \\
         29  & -0.352 & -0.023 & -0.361 & IIA & 58.0 &  1 \\
         30  & -0.263 & -0.341 &  0.003 &  IA & 23.7 & -1 \\
         31  & -0.270 &  0.424 &  0.003 &  IB & 79.2 &  1 \\
         32  & -0.263 &  0.341 & -0.003 &  IA & 23.7 &  1 \\
          & & & & & & \\
        \hline
        \end{tabular}
        \end{center}
        \label{table:wp}
    \end{table*}
    
    \subsection{Robustness of altermagnetic phase, effect of spin-orbit coupling and bulk Weyl points}
    In Sec. III(C) of the main manuscript, we mentioned that \textcolor{black}{GdAlSi retains the altermagnetic band structure even in the absence of strong correlations arising due to $f$-electrons of Gd and that the} incorporation of spin-orbit coupling (SOC) in the first-principles calculation did not have much effect on the electronic band structure of GdAlSi. Here we show the electronic band structure of GdAlSi for AFM-I magnetic configuration in the absence of any on-site Coulomb correlations taken into account (U = 0),  without and with SOC, as shown in Figs. \ref{fig:figs4}(a,b). We have also calculated the electronic band structure of GdAlSi in paramagnetic phase (Gd $f$-orbitals frozen in core) both without and with SOC as shown in Figs. \ref{fig:figs4}(c,d). The band structure of AFM-I magnetic configuration with on-site Couloumb correlations taken into account (U = 7.0 eV), both without and with SOC, has been shown in Figs. 2(d,e) of the main manuscript. Without Hubbard U, the band structure in Fig. \ref{fig:figs4}(a) still shows momentum-dependent spin-splitting with similar features as seen in Fig. 2(d) of the main manuscript. The most notable difference is the presence of Gd-$f$ conduction bands coming near the Fermi level. These bands move away from the Fermi level with the inclusion of U which is expected due to further localization of Gd-$f$ electrons. The band structure with SOC shown in Fig. \ref{fig:figs4}(b) is similar to that of without SOC since Gd is an effective $s$-state ion as explained in the main manuscript. The minor differences observed are further explained below. In the absence of SOC and any magnetic order, all the electronic bands are doubly degenerate due to the presence of $\mathcal{T}$ symmetry (Kramers' degeneracy) as shown in Fig. \ref{fig:figs4}(c). Upon inclusion of SOC effect, this degeneracy is lifted as a consequence of the non-centrosymmetric crystal structure of GdAlSi, which induces Rashba/Dresselhaus type of spin-splitting as shown in Fig. \ref{fig:figs4}(d). But this splitting is much less compared to the altermagnetic band splitting when considering AFM-I magnetic order without SOC as discussed at length in the main manuscript (also see Fig. 2(d) of the main manuscript). When SOC effect is included, most notably, the spin-degeneracy along the $\Gamma-\text{X}-\text{P}-\text{U}-\text{Z}$ direction is lifted, whereas this effect is masked in the other directions as has been shown in Fig. 2(e) of the main manuscript.
    The total magnetic moment of individual Gd atoms in the AFM-I configuration remains equal in magnitude but parallel/antiparallel to the crystalline $c$-axis. This is also seen if the spin components along the $z$-axis (crystalline $c$-axis), referred to as $\sigma_z$, is projected on the SOC band structure as mentioned in the main manuscript. The $\sigma_z$ projected SOC band structure along with the total Berry curvature at the Fermi level is shown in Fig. \ref{berry}. Since SOC perturbs the bands only slightly, the \textit{alternating} nature of $\sigma_z$ projections remain intact (see Fig. \ref{berry}(a)) when compared with the spin-polarized band structure shown in Fig. 2(d) of the main manuscript. This \textit{alternating} nature is further reflected in the total Berry curvature as shown in Fig. \ref{berry}(b). Due to avoided crossings along the $\Gamma-\Sigma-\text{N}-\Sigma_1-\text{Z}$ and $\Gamma-\Sigma^{\prime}-\text{N}'-\Sigma_1^{\prime}-\text{Z}$ paths, the bands along these two paths have equally large but opposite contributions to the total Berry curvature. This can lead to highly direction-dependent anomalous Hall and anomalous spin Hall effect in GdAlSi which can have practical applications in devices proposed in Sec. III(G) of the main manuscript. 
    Another consequence of SOC is the emergence of bulk Weyl points (WPs) with definite chirality as shown in Fig. 3 of the main manuscript. The details about the location of bulk WPs in energy and $k$-space are presented in Table \ref{table:wp}.

    \subsection{Role of bulk altermagnetism in inducing spin-splitting on (001) surface}

    In the main manuscript, it was explained that the (001) surface of GdAlSi possesses a different symmetry than the bulk which results in an asymmetry in the 2D FS. We further noted that non-magnetic compounds in $I4_{1}md$ space group also showed this 2D FS asymmetry. Here, we show the 2D FS of GdAlSi for paramagnetic configuration without SOC and with SOC in Fig. \ref{fig:figs6}(a) and \ref{fig:figs6}(b) respectively. The 2D FS is again asymmetric as expected due to breaking of $C_{4z}$, $\mathcal{M}_{xy}$ and $\mathcal{M}_{x\overline{y}}$ symmetries in $mm2$ ($mm21'$ with SOC). The surface bands in Fig. \ref{fig:figs6}(a) are doubly-degenerate due to presence of $\mathcal{PT}$ symmetry in momentum space without SOC. Although the FS appears to have a four-fold symmetry, the asymmetry is noticeable near the $\overline{\text{X}}$ and $\overline{\text{Y}}$ points. Hence, it actually has a two-fold symmetry. In Fig. \ref{fig:figs6}(b), the two-fold symmetry becomes much more apparent since the large surface pockets forming a diamond-like structure disappear and open Fermi arcs appear joining WPs, similar to the 2D FS in Fig. 3(b) of the main manuscript. Further, the central pocket around $\overline{\Gamma}$ is finely split. The noticeable difference here is that the splitting of the Fermi arcs and the surface pockets are minor. Since $\mathcal{P}$ symmetry is absent in real-space, the surface bands become spin-split in the presence of SOC as a consequence of Rashba/Dresselhaus effects. When AFM-I order is considered without SOC, the 2D FS is spin-polarized as shown in Figs. \ref{fig:figs6}(c) and \ref{fig:figs6}(d). It is seen that the spin-up and spin-down 2D FS are quite different and the Fermi arcs do not \textit{alternate}, although the surface-projected bulk Fermi pockets can be seen to \textit{alternate} as in Fig. \ref{spin_FS}. The 2D FS asymmetry can again be seen in both Figs. \ref{fig:figs6}(c) and \ref{fig:figs6}(d). Upon inclusion of SOC, these two separate spin-polarizations hybridize and generate the 2D FS shown here in Fig. \ref{fig:figs6}(e) and Fig. 3(b) of the main manuscript. As noted before, the spin-splitting is large compared to Fig. \ref{fig:figs6}(b), especially for the central pocket. Figure \ref{fig:figs6}(f) highlights a quadrant of the surface BZ showing the Fermi arc connectivity of the surface projections of the type-I WPs. In the bulk, alternating band structure was seen along the $\Gamma-\Sigma-\text{N}-\Sigma_1-\text{Z}$ and $\Gamma-\Sigma^{\prime}-\text{N}'-\Sigma_1^{\prime}-\text{Z}$ paths. These paths get projected to $\overline{\Gamma}-\overline{\text{X}}$ and $\overline{\Gamma}-\overline{\text{Y}}$ paths respectively on the (001) surface. The surface dispersion along these paths is shown in Figs. \ref{fig:figs6}(g-i). Without SOC, the surface band structure is spin-polarized. From Figs. \ref{fig:figs6}(g) and \ref{fig:figs6}(h), it is seen that the surface bands do not \textit{alternate} about $\overline{\Gamma}$ although the surface projections of the bulk pockets \textit{alternate} as expected. Nevertheless, the bulk altermagnetic order induces spin-polarization of surface bands leading to large spin-splitting. This is more evident in Fig. \ref{fig:figs6}(i) where the spin-up and spin-down surface bands hybridize in the presence of SOC. Thus, GdAlSi shows an interplay of bulk altermagnetism and non-trivial topology on the surface.

    \section{ARPES results} \label{expt}
    
    \subsection{Surface Band Dispersion}

    \begin{figure}[!htb]
        \centering
        \includegraphics[width=0.7\linewidth]{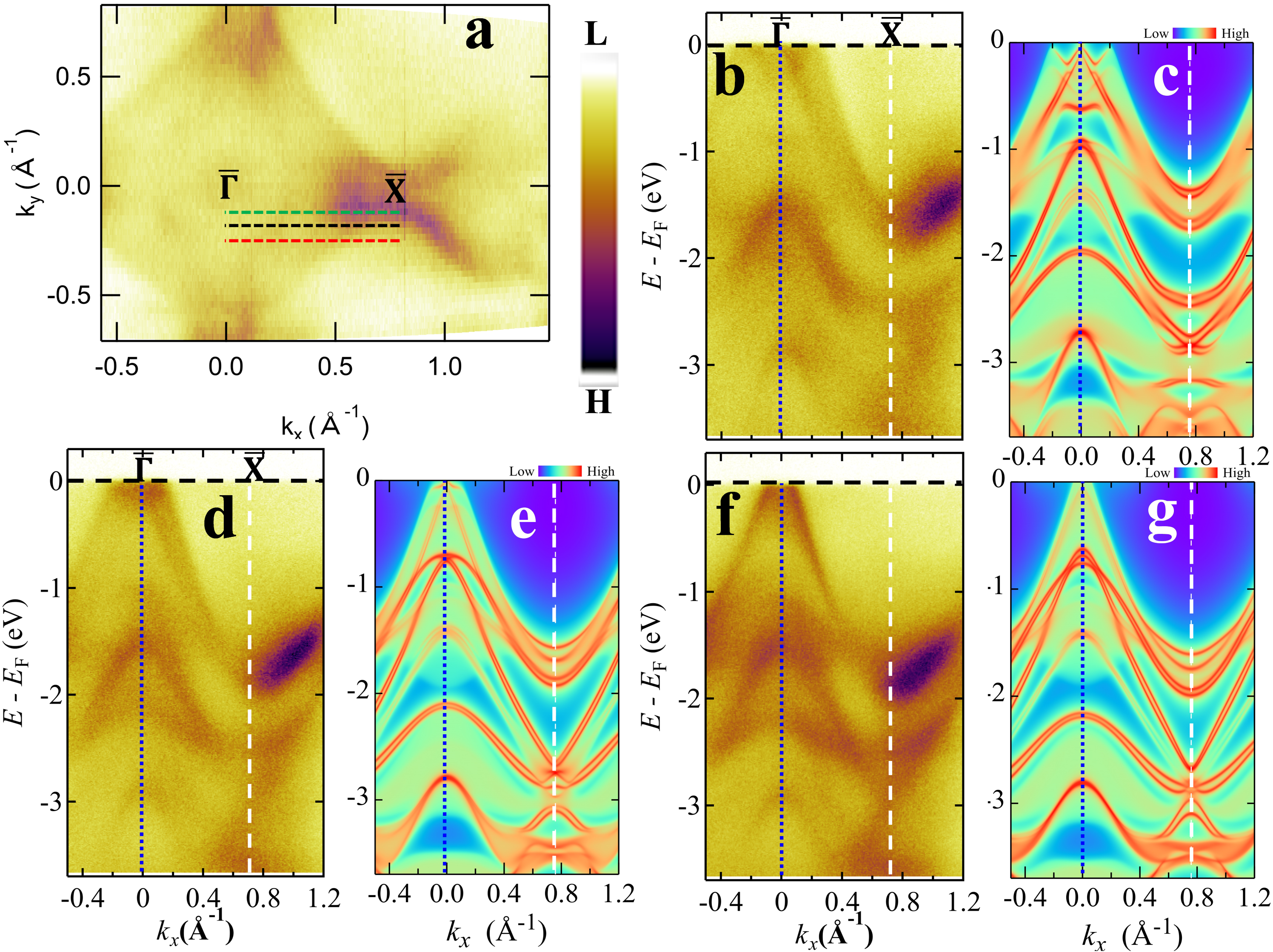}
        \caption{(a) For the (001) surface of GdAlSi, symmetrized measured Fermi surface using a photon energy of 50 eV. Energy-momentum cuts were performed at three different \textit{k$_y$} positions along $\overline{\Gamma}$-$\overline{\text{X}}$, indicated by green, black, and red dashed lines. (b, d, f)  Corresponding ARPES band structures along the colored dashed lines (green, black, and red) respectively. (c, e, g) Calculated band structures along the same cuts. The vertical blue dotted and white dashed line locate $\overline{\Gamma}$ and $\overline{\text{X}}$ points respectively.}
        \label{fig:figss7}
    \end{figure}

    Figure \ref{fig:figss7} illustrates the Fermi surface of GdAlSi, showing a diamond-shaped structure  featuring a symmetrical four-fold pattern along the (001) direction, measured using a photon energy of 50 eV. We study energy-momentum cuts of band dispersion taken at three distinct $k_y$ positions (cuts pass through the Fermi arcs as predicted by the calculated FS) aligned with $\overline{\Gamma}$-$\overline{\text{X}}$ line. These positions are denoted by the green (lies on $\sim$ $k_y = 0.17$\AA$^{-1}$), black (lies on $\sim$ $k_y = 0.21$\AA$^{-1}$), and red (lies on $\sim$ $k_y = 0.24$\AA$^{-1}$) dashed lines as shown in Fig. \ref{fig:figss7}(a). The corresponding ARPES band dispersion along the specified dashed lines (green, black, and red) are presented in Fig. \ref{fig:figss7}(b, d, f), respectively. For a direct comparison, band structures computed along the same cuts are shown in Fig. \ref{fig:figss7}(c, e, g), respectively. ARPES data reveal several interesting features, such as (i) pronounced linear band dispersion around $\overline{\Gamma}$ in close proximity to $E_F$, (ii) a strong, prominent band feature at $\sim$ -1.5 eV near the edge of the bulk-energy gap at $\overline{\text{X}}$ (iii) two downward parabolic band dispersions at energies $\sim$ -1.5 eV and -2.5 eV, respectively near $\overline{\Gamma}$ separated by the bulk energy gap (iv) A subtle downward parabolic-like band dispersion at
    $\sim$ -3.0 eV at $\overline{\Gamma}$.These significant features, particularly (i-iii), exhibit minimal variations in terms of energy and position across different energy-momentum cuts. The experimental data also clearly depict bulk energy gaps that align with the features in simulated band structure.  
    This results in a good qualitative agreement between the experiment and theory.

    \subsection{Fermi arc-like feature}

    \begin{figure*}[!htb]
        \centering
        \includegraphics[width=0.9\linewidth]{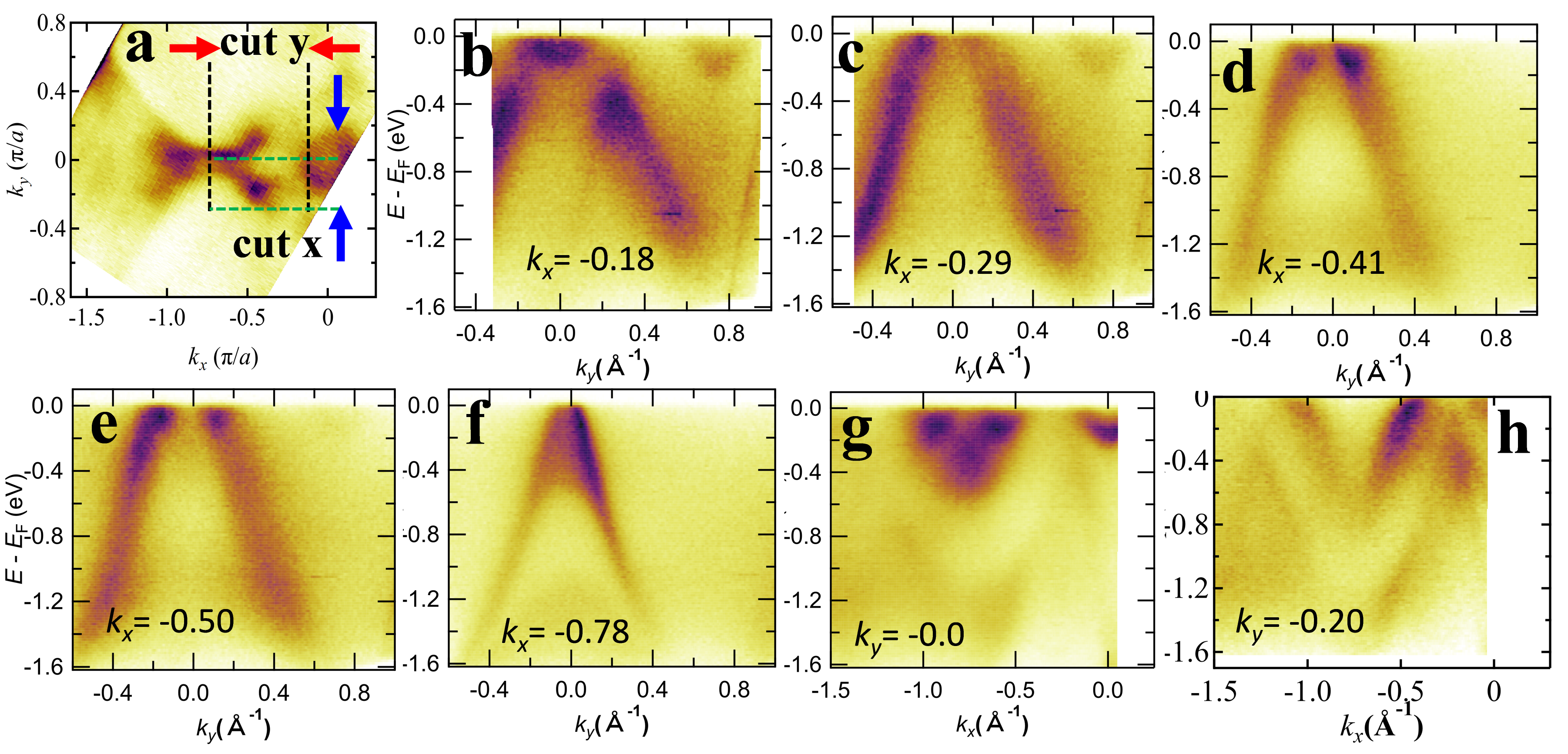}
        \caption{For GdAlSi (a) low energy (50 eV) ARPES un-symmetrized FS along the (001) direction with energy momentum cut along the type-I and type-II Fermi arcs as observed in calculated FS. Black dashed lines represent the cut along the $\overline{\Gamma}-\overline{\text{Y}}$ direction, while green dashed lines represent the cut parallel to the $\overline{\Gamma}-\overline{\text{X}}$ direction. Band dispersion along (b-f) cut y and (g-h) cut x, as shown in Fig. (a).}
        \label{fig:fig6s}
    \end{figure*}
    
    Our ab initio calculated Fermi surface indicates that type-II Fermi arcs are situated along the $\overline{\Gamma}-\overline{\text{X}}$ direction, while the type-I Fermi arcs are oriented along the $\overline{\Gamma}-\overline{\text{M}}$ direction as shown in the Fig. \ref{fig:figs6}. To investigate the Fermi arc-like characteristics, we analyze energy-momentum cuts along both the horizontal and vertical directions to the $\overline{\Gamma}-\overline{\text{X}}$ line. Figures \ref{fig:fig6s}(b-f) present energy-momentum cuts along the line perpendicular to the $\overline{\Gamma}-\overline{\text{X}}$ path (cut y) at various $k_x$ positions, while Figs. \ref{fig:fig6s}(g-h) depict the horizontal cuts along the same path (cut x). The corresponding paths are shown in Fig. \ref{fig:fig6s}(a). ARPES measurements reveal mirror symmetric strong band dispersion features ($\sim$ $k_y= 0.3$\AA$^{-1}$) around the electron pocket ($\overline{\Gamma}$) for the cut y, as shown in Fig. \ref{fig:fig6s}(b). Mirror symmetric linear band dispersion features with negligible $k_z$ dispersion become more prominent for the cuts passing through type-I/II Fermi arcs as shown in Fig. \ref{fig:fig6s}(c-e). As observed in Fig. \ref{fig:fig6s}(d-e) along the cut y across the Fermi arcs (for $\sim$ $k_x= -0.4$\AA$^{-1}$ plane), two counter-propagating edge modes-like signatures are visible, which could be associated with opposite chiral charges ($\pm$1).
    By comparing these experimental signatures obtained from the VUV ARPES results, with those of the predicted outcomes from our calculations, we offer an interpretation of the probable locations and existence of the predicted Weyl Fermions and potential Weyl Fermi arcs. Interestingly, a club-like feature with minimal $k_z$ dispersion is also observed in Fig. \ref{fig:fig6s}(g) for the cut x at $\sim$ $k_y= 0.0$.
    
    \putbib[sm]

\end{bibunit}

\end{document}